\documentclass[twocolumn,showpacs,amsmath,amssymb]{revtex4}

\usepackage{epsfig,amscd}
\usepackage{dcolumn}
\usepackage{bm}

\newcommand{\bs}[1]{
             \mathbf{#1}}
\newcommand{\fr}[1]{
             \frac{#1}}
\newcommand{\no}{\nonumber}
\newcommand{\be}{\begin{equation}}
\newcommand{\ee}{\end{equation}}
\newcommand{\bea}{\begin{eqnarray}}
\newcommand{\eea}{\end{eqnarray}}
\newcommand{\Gr}{G(\bs{r}|\bs{r}^\prime ; t)}
\newcommand{\Grb}{G(\bs{r}^\prime|\bs{r} ; t)}
\newcommand{\Gx}{G(x,y|x^\prime,y^\prime ; t)}

\newcommand{\ax}{(x,y)}

\newcommand{\axp}{(x^\prime,y^\prime)}

\newcommand{\lplc}{\nabla^2}

\newcommand{\ihat}{\hat{\mbox{\boldmath{$i$}}}}
\newcommand{\jhat}{\hat{\mbox{\boldmath{$j$}}}}
\newcommand{\khat}{\hat{\mbox{\boldmath{$k$}}}}

\newcommand{\rvek}{\mbox {\boldmath $r$}}

\newcommand{\di}{\displaystyle}
\newcommand{\difrac}[2]{\frac{\di {#1}}{\di {#2}}}
\newcommand{\didrv}[2]{\frac{\di\partial{#1}}{\di\partial{#2}}}

\newcommand{\sxth}{\difrac{1}{6}}

\newcommand{\eqn}[1]{(\ref{#1})}

\begin{document}

\title{NMR for Equilateral Triangular Geometry
under Conditions of Surface Relaxivity - Analytical and Random Walk Solution}

\author{A. Hiorth}
\email{ah@rf.no}
\affiliation{RF-Rogaland Research, P. O. Box 8046, N-4068 Stavanger,
  Norway}
\author{U. H. a Lad}
\affiliation{RF-Rogaland Research, P. O. Box 8046, N-4068 Stavanger,
  Norway}
\author{J. Finjord}
\affiliation{University of Stavanger, N-4036 Stavanger,Norway}
\author{S. M. Skj\ae veland}
\affiliation{University of Stavanger, N-4036 Stavanger,Norway}

\date{\today}
\begin{abstract}
We consider analytical and numerical solution of NMR relaxation
under the condition of surface relaxation in an equilateral
triangular geometry. We present an analytical expression for the
Green's function in this geometry. We calculate the transverse
magnetic relaxation without magnetic gradients present, single-phase, both
analytically and numerically. There is a very good match between
the analytical and numerical results. We also show that the magnetic
signal from an equilateral triangular geometry is qualitatively
different from the known solution: plate, cylinder and sphere, in the
case of a nonuniform initial magnetization. Non uniform magnetization
close to the sharp corners makes the magnetic signal very fast multi
exponential. This type of initial configuration fits qualitatively
with the experimental results by Song et al. \cite{song}.
It should also be noted that the solution presented here can be used
to describe absorption of a chemical substance in an equilateral
triangular geometry (for a stationary fluid). 
\end{abstract}


\maketitle

\begin{widetext}
\section{Introduction}
\label{}
In the oil industry there has been a great interest in using NMR as a
tool for improved reservoir characterizing. NMR can be used as an in
situ tool for measuring oil and brine content in saturated porous
rocks\cite{freedman,heaton,hurlimann,hurlimann2}. The distribution of
oil and water in the porous rock is also very important. If a
phase is in contact with pore wall, it will
experience enhanced relaxation\cite{brownstein1,brownstein2}. One way
of observing this is by studying how the oil signal changes when the
core is aged, see \cite{howard,zhang} and the references therein. The
distribution of oil and water in a porous rock is mainly determined by
the geometry of the pore space and the chemical composition of oil,
brine and rock. The NMR signal from a porous rock will then be a strong
function of the specific pore geometry and the wetting status of the
rock. Experiments on porous rocks are well suited for showing an effect of
enhanced surface relaxation due to wettability change. However, by
study simpler systems, more information can be gained in order
to understand how chemical properties  will affect the magnetic signal. 

By assuming a specific geometry for the porespace more
information from the NMR signal can be obtained. Cylindrical and
spherical geometries can be used when wettability is not a topic. However most
reservoirs are mixed wet and one need geometries containing sharp
corners.  This realization is the starting point 
for this work. One of the simplest geometries which allows for more than 
one phase to form a stable configuration is an equilateral triangular geometry.  
A great deal of
attention have been devoted to triangles in order to understand the
multi phase behavior of porous rocks\cite{radke,morrow1,morrow2}.
Triangles are the basic building
blocks in pore network models\cite{blunt} and also bundle of
equilateral triangular tubes models have been studied extensively
\cite{sorbie,helland}. Attempt of using a bundle of equilateral
triangular tubes in interpreting NMR signal from porous rocks have
also been done\cite{al-mahrooqi}.

As always with NMR experiments the key challenge is to get a proper interpretation
of the magnetic signal. The crucial step in order to gain precise
knowledge from NMR signal  
is to have a good theoretical description of the magnetic signal. 
This work will only be concerned with magnetic signal from single phases, but it forms a 
necessary basis in developing numerical code for 
magnetic signal from triangles containing
multi phases. The numerical solution presented in this work can be
extended to account for more than one phase inside the rock and 
analytical single phase results is necessary in order to calibrate
the numerical algorithm. The multi phase numerical code can then be used to
study how wettability will influence the magnetic signal in a well
defined geometry. Experiments on glass triangular tubes are in
progress at our group and the analytical solution presented here  
will be used to interpret the experimental result, this will be
considered in a different work. 

The higher modes presented here are usually neglected, but have been
shown to give valuable information of pore sizes
\cite{song,marinelli}. In order to fully characterize a geometry one
need information from higher modes. The lowest mode is dependent on
the surface relaxivity and proportional to the internal surface area
divided by the total volume of the pore ($S/V$), the higher modes are
independent of the surface relaxivity and proportional to $(S/V)^2$.
A clever way of enhancing the weight of the higher modes is by
creating a non uniform initial magnetization, as described by \cite{song}.

The outline of the papers is as follows: in section \ref{sec:th} we
present the theory for magnetic relaxation in a confined space. In
section \ref{sec:green}, we calculate the Green's function for an
equilateral triangular geometry and in section \ref{sec:mag} we
calculate the magnetization for different initial configurations.
Then we present details of the random walk solution in section \ref{sec:rw} and
finally conclusion and discussion. 

\section{Theory}\label{sec:th}
The equilateral triangular geometry is a true two dimensional system,
contrary to plate, cylinder and sphere geometry, first considered by
Brownstein and Tarr\cite{brownstein2}. This problem is much harder to
solve, the technical reason for this is that it is not possible to
choose a coordinate system where the axes are parallel to the sides of
the triangle. Fortunately the calculation greatly simplifies because
of a series of recent papers by Mccartin, which solves the diffusion
equation in an equilateral triangular
geometry\cite{mccartin1,mccartin2,mccartin3}. The magnetization as a
function of time is given by the following equations\cite{brownstein1,brownstein2}:
\bea
&&\didrv{M(\bs{r},t)}{t}=D\lplc
M(\bs{r},t)-\frac{M(\bs{r},t)}{{T_2}_b}\, .
\label{diffm}
\eea
${T_2}_b$ is the bulk relaxation  and $D$ the diffusion constant. At
the surface, $\Sigma$, we have the following boundary condition:
\bea
&&D\hat{\bs{n}}\cdot\nabla M(\bs{r},t) +\rho
M(\bs{r},t)|_{\bs{r}=\Sigma}=0\, ,
\label{mbound}
\eea
where $\hat{\bs{n}}$ is the outward normal and $\rho$ the surface
relaxivity. Equation (\ref{diffm}) (with $1/{T_2}_b = 0 $) 
and (\ref{mbound}) are identical to the equations governing absorption
of a chemical substance on a surface, in the case of a stationary fluid. 
The magnetization of the sample is :
\bea
M(t)=\int_\Delta d\bs{r}M(\bs{r},t)\, ,\label{mag}
\eea
where the integral is taken over the triangular domain.
By introducing  the Green's function we can write the magnetization as:
\begin{equation}
M(\bs{r},t)=e^{-t/{T_2}_b}\int_\Delta
d\bs{r}^\prime\,\rho(\bs{r}^\prime)\Grb\, ,
\label{cpgm}
\end{equation}
where $\rho(\bs{r}^\prime)$ is the initial spin density.
$\Grb$ is the Green's
function or the propagator. It is defined
as the probability for a particle at position $\bs{r}^\prime$ at time 0 to
diffuse to point $\bs{r}$ during a time $t$. The propagator satisfies
the diffusion equation at the interior of the pore space:
\bea
&&\frac{\partial\Gr}{\partial t}-D\nabla^2\Gr=0;\;\text{and}\no\\
&&\Gr|_{t=0}=\delta(\bs{r}-\bs{r}^\prime)\, ,\label{gr}
\eea
where $D$ is the diffusion constant. The boundary condition at the
surface $\Sigma$ :
\be
D \hat{\bs{n}}\cdot\nabla\Gr+\rho\,\Gr|_{\bs{r}=\Sigma}=0\, .
\label{grb}
\ee
\begin{figure}[t]
\begin{center}
   \epsfig{file=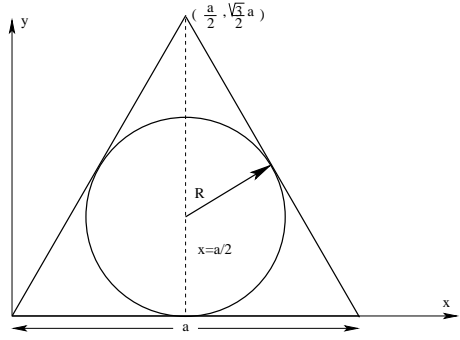,width=6cm}
\caption{Equilateral triangle of side length $a$ and inscribed radius $R$}
\label{fig:eq1}
\end{center}
\end{figure}

\section{The Green's function in an equilateral triangular geometry}\label{sec:green}
In this section we will discuss relevant eigenfunctions for situations where there is no gradients present, for the general case see Appendix \ref{appa}.
By using the standard eigenfunction expansion of the propagator:
\be
\Gr=\sum_{i=0}^\infty\phi_i(\bs{r})\phi_i(\bs{r}^\prime)e^{-t/T_i}\, ,
\ee
where $\{\phi_i\}$ are an orthonormal set of eigenfunctions with
corresponding eigenvalues $T_i$.
From equation (\ref{gr}) and (\ref{grb}) it then follows:
\bea
&&D\nabla^2\phi_i(\bs{r})=-\fr{1}{T_i}\phi_i(\bs{r})\;\;\text{and}\no\\
&&D\hat{\bs{n}}\cdot\nabla\phi_i(\bs{r})+\rho\,\phi(\bs{r})|_{\bs{r}=\Sigma}=0
\, .
\label{phi}
\eea
This equation needs to be solved in an equilateral triangle, shown in
Figure \ref{fig:eq1}. The orthonormal set of eigenfunctions can be
divided in a set of eigenfunctions symmetric and antisymmetric about
the line $x=a/2$, i.e.
$\{\phi\}=\{\phi^s,\phi^a\}=\{T^s/N^s,T^a/N^a\}$. $\{T^s,T^a\}$ is the
set of orthogonal eigenfunctions and $N^{s,a}$ the normalization. When the
magnetization is given by equation (\ref{cpgm}), the only relevant
modes are the symmetric diagonal modes (for the full solution see
Appendix \ref{appa}) :
\bea
&T_{ii}^s(x,y)&=\cos\left[\fr{2\pi\mu}{3R}y-\delta_2\right]
+2\cos\left[\fr{\pi\mu}{\sqrt{3}R}(\sqrt{3}R-x)\right]
\cos\left[\fr{\pi\mu}{3R}(y-3R)+\delta_2\right]\no\\
&T_{ii}^a(x,y)&=0\no\\
&{N^s_{ii}}^2&=\int_0^{3R}\int_{y/\sqrt{3}}^{-y/\sqrt{3}+2\sqrt{3}R}dxdy\,
T^{s}_{ii}(x,y)\,T^{s}_{ii}(x,y)\no\\
&&=\frac{9\sqrt{3}R^2}{16(\mu\pi)^2}\big\{8(1+\mu^2\pi^2)-7 \cos [2 {\delta_2}]
-8\cos [2 \mu \pi ]\no
-\cos [2 {\delta_2}-4 \mu \pi ]
+8 \cos [2 {\delta_2}-2 \mu \pi]\no\\&&\quad\quad\quad\quad
+ 4\mu\pi\sin [2 {\delta_2}]
-16 \mu \pi  \sin [2 {\delta_2}-2\mu \pi ]\big\}\label{nii}
\eea
and the eigenvalues :
\be
T_{ii}^{-1}=\fr{4D}{9}\left(\fr{\pi\mu}{R}\right)^2\, .\label{diag}
\ee
$\mu$ and $\delta_2$ are determined from the boundary condition
(\ref{phi}) and it give rise to the following transcendental equation:
\bea
&&\left[1-\frac{9}{2}
\left(\frac{\gamma}{\pi\mu}\right)^2\right]\tan\pi\mu=\frac{9\gamma}{2\pi\mu}\;,
\qquad\mu\in <i,1+i]\no\\
&&\delta_2=\tan^{-1}\frac{3\gamma}{2\pi\mu}\label{tranc}\, 
\qquad\text{and}\qquad\gamma=\frac{\rho R}{D}
\eea
A reduced from of the propagator, when the magnetic signal 
is given by equation  (\ref{cpgm}) is then:
\bea
&&\Gx=\sum_{i=0}^{\infty}\frac{T^s_{ii}\ax T^s_ii\axp
}{{N^s_{ii}}^2}e^{-t/T_{ii}}\label{green}
\eea
This equation can now be used to calculate the magnetic signal  
from an equilateral triangular pore given an initial spin density.

\section{Magnetic Signal from an Equilateral Triangle}\label{sec:mag}

From equation
(\ref{mag}), (\ref{cpgm}) and (\ref{green}) we find :
\bea
&M(t)&=e^{-t/{T_2}_b}\sum_{i=0}^\infty\int\int d\bs{r}d\bs{r}^\prime \rho(\bs{r}^\prime)
\phi^s_{i}(\bs{r})\phi^s_{i}(\bs{r}^\prime) e^{-t/T_{ii}}\no\\
&&= e^{-t/{T_2}_b}\frac{9\sqrt{3}R^2}{2}\sum_{i=0}^\infty
\frac{e^{-t/T_{ii}}}{N_{ii}^2(\mu\pi)^2}\int d\bs{r}^\prime 
\rho(\bs{r}^\prime)T^s_{ii}(\bs{r}^\prime)
\left(\cos[\delta_2] - \cos[\delta_2 - 2 \mu\pi] +
              2 \mu\pi \sin[\delta_2]\right)
\eea
We need to know the initial excited spin density. Usually it is assumed that the 
initial spin density is uniform as in the work by Brownstein and Tarr\cite{brownstein2}. 
However, recently Song et al.\cite{song} have shown during a series of papers that it
is possible to create a nonuniform initial magnetization. The physical explanation
for this is due to susceptibility differences between water and rock\cite{song}. 
There are then four interesting cases, shown in table \ref{tabd}.
\begin{table}[t]
\begin{center}
\caption{Different initial spin densities}
\label{tabd}
\begin{tabular}{|c| c|}\hline\hline
 Type            &  $\rho(x,y)$ \\ \hline
 uniform         &  $1/(3\sqrt{3}R^2)$ \\ \hline
 center          &  $\delta\left(x-\sqrt{3}R\right)\,\delta\left(y-R\right)$ \\ \hline
 side            &  $\delta\left(x-\sqrt{3}R\right)\,\delta\left(y-R/32\right)$ \\ \hline
 corner          &  $\delta\left(x-\sqrt{3}R\right)\,\delta\left(y-7R/32\right)$ \\ \hline\hline
\end{tabular}
\end{center}
\end{table} 

For the side and corner densities we have chosen points close to the
surface, which lies at one of the symmetry line of the triangle. Then
we only need to consider one corner(side) and the result will be
equivalent for the other corners(sides).
\begin{figure*}[t]
\begin{center}
$\begin{array}{c c}
\epsfxsize=7.2cm
\epsffile{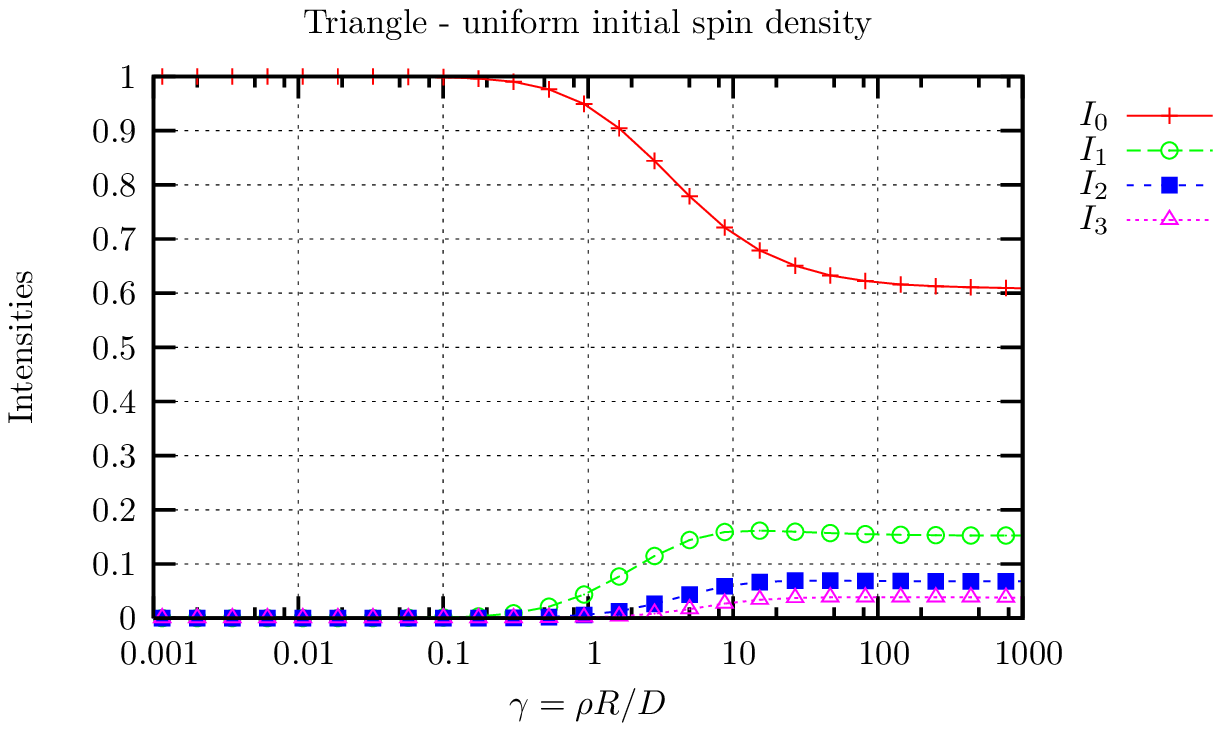} &
    \epsfxsize=7.2cm
    \epsffile{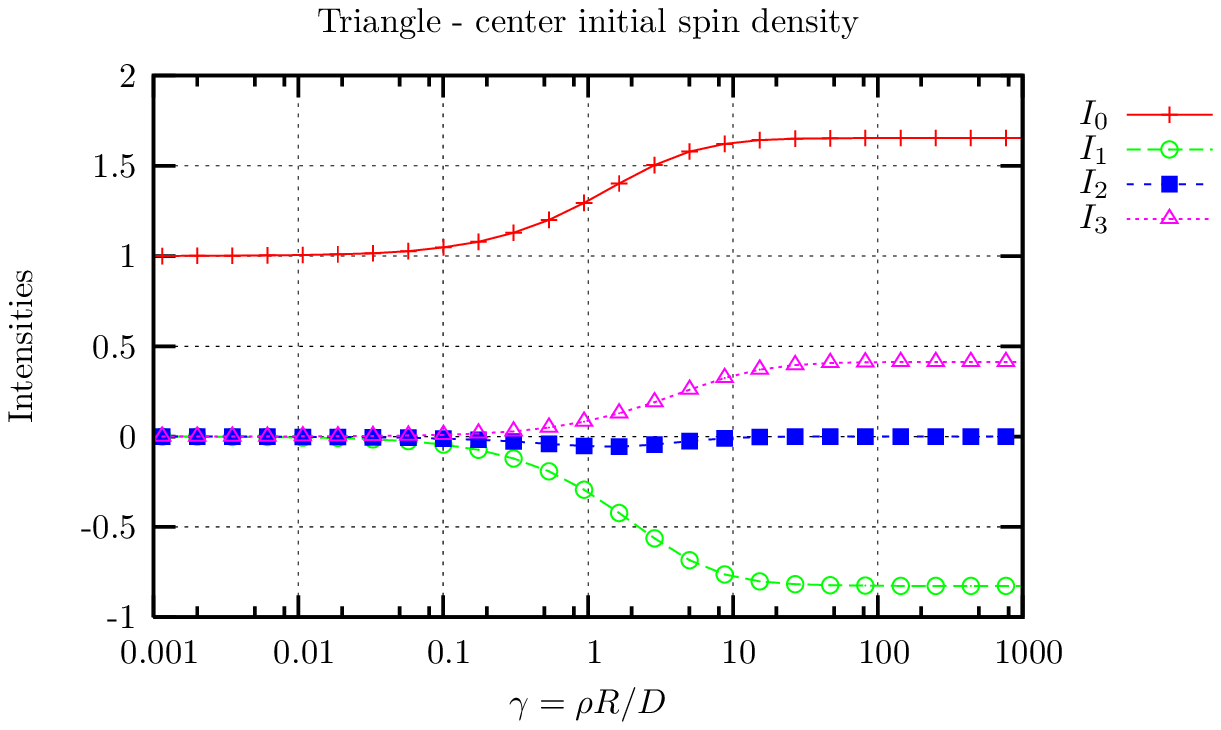} \\
\epsfxsize=7.2cm
\epsffile{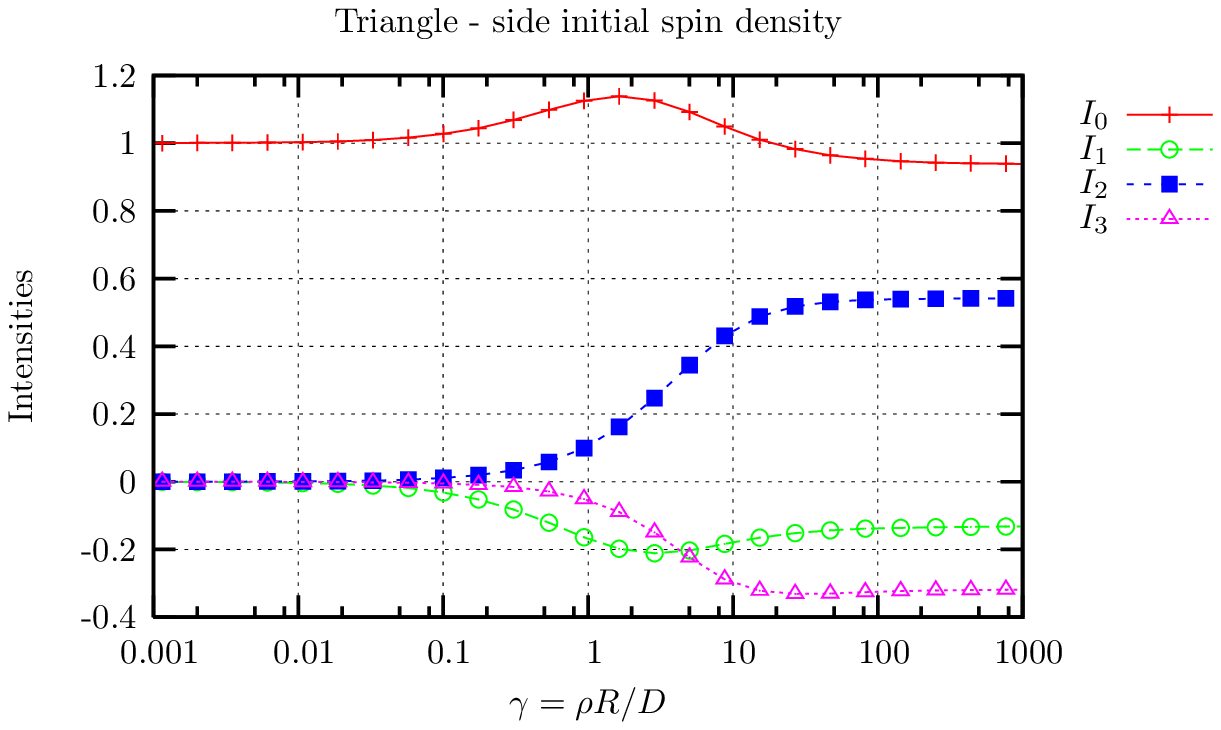} &
    \epsfxsize=7.2cm
    \epsffile{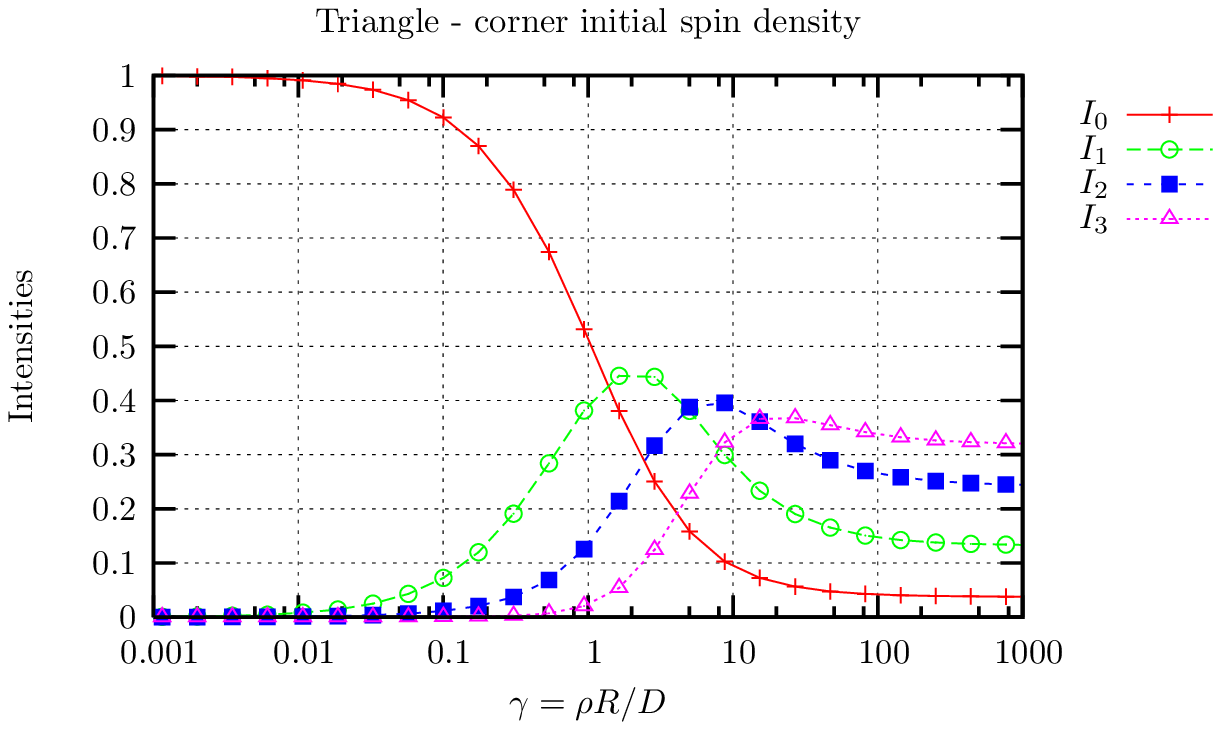} 
\end{array}$
\end{center}
\caption{Upper-Left: the first four intensities for uniform initial spin
  density ($I_0\to0.61,\,I_1\to0.15,\, I_2\to0.07,\, I_3\to 0.04$). 
 Upper-right: the first four modes for all spins initially at the
  center of the triangle ($I_0\to 1.65,\,I_1\to -0.83,\, I_2\to 0,\,
  I_3\to 0.41$).
 Lower-left: the first four modes for all spins initially at the
  side of the triangle ($I_0\to 0.94,\,I_1\to -0.13,\, I_2\to 0.54,\,
  I_3\to -0.32$).
 Lower-right: the first four modes for all spins initially at the
  corner of the triangle ($I_0\to 0.04,\,I_1\to 0.13,\, I_2\to 0.24,\, I_3\to 0.32$).}
\label{fig:In}
\end{figure*}

\bea
&M^{\text{uniform}}(t)
&=e^{-t/{T_2}_b}\sum_{i=0}^{\infty}\frac{27\sqrt{3}R^2}{4(\mu\pi)^4N_{ii}^2}\,
e^{-t/T_{ii}}
\left(\cos[\delta_2]- \cos [\delta_2-2 \mu \pi ]+2 \mu \pi \sin [\delta_2]\right)^2\\
&M^{\text{center}}(t)
&=e^{-t/{T_2}_b}\sum_{i=0}^{\infty}\frac{27\sqrt{3}R^2}{2(\mu\pi)^2N_{ii}^2}\,
e^{-t/T_{ii}}
\cos[\delta_2-2\mu\pi/3]
\left(\cos[\delta_2]- \cos [\delta_2-2 \mu \pi ]+2 \mu \pi \sin [\delta_2]\right)\\
&M^{\text{side}}(t)
&=e^{-t/{T_2}_b}\sum_{i=0}^{\infty}\frac{9\sqrt{3}R^2}{2(\mu\pi)^2N_{ii}^2}\,
e^{-t/T_{ii}}\left(2\cos[\delta_2-7\mu\pi/8]+\cos[\delta_2-\mu\pi/4]\right)\no\\
&&\quad\times
\left(\cos[\delta_2]- \cos [\delta_2-2 \mu \pi ]+2 \mu \pi \sin [\delta_2]\right)\\
&M^{\text{corner}}(t)
&=e^{-t/{T_2}_b}\sum_{i=0}^{\infty}\frac{9\sqrt{3}R^2}{2(\mu\pi)^2N_{ii}^2}\,
e^{-t/T_{ii}}\left(\cos[\delta_2-7\mu\pi/4]+2\cos[\delta_2-\mu\pi/8]\right)\no\\
&&\quad
\times
\left(\cos[\delta_2]- \cos [\delta_2-2 \mu \pi ]+2 \mu \pi \sin[\delta_2]\right)\, ,
\eea
where $N_{ii}^2$ is given in equation (\ref{nii}).

We have compared the analytical solution for the equilateral triangle
with the solution for plate, cylinder and sphere, with different
initial spin densities. Initial spin density close to a corner clearly
does not exists for those geometries. We do not show the result for
plate, cylinder and  sphere here, but they behave
qualitatively in the same manner. When approaching the slow diffusion
limit (SDL, see next subsection) the weight of
the higher modes increases compared to the uniform case, but the
lowest mode always gives the highest contribution to the magnetic
signal, see Figure \ref{fig:In}. From Figure \ref{fig:In} it is evident
that initial spin density close to a corner is qualitatively
different from uniform, center and side. Sharp corners and initial
spin density close to the corners make the magnetic signal  multi
exponential very fast, even for $\gamma=1$. For $\gamma=4$ the lowest
mode is the {\it least} dominant one.

In order for the lowest mode
to be dominant the spin density for all times must be more or less
uniform. In the non uniform cases, the spins lose coherence due to
surface effects before the magnetization is uniform. This has the
effect that non uniform initial spin density in general makes the
higher modes more dominant. In order to suppress the lowest mode, the
surface effects must be felt before the spins manage to diffuse over
the triangular domain to give a uniform magnetization. Clearly,
corners will be very efficient  of suppressing the lowest mode. This
also give a natural explanation as to why this effect is not so prominent
for plate, cylinder and sphere.  
 
\subsection{Fast Diffusion and Slow Diffusion Limit}

There are two different time scales of interest. The relaxation time which is
dependent on the surface relaxivity, $\tau_\rho\sim R/\rho$ and the
diffusion time which is dependent on the diffusion constant
 $\tau_D\sim R^2/D$. If $\tau_D<<\tau_\rho$ we are in
the fast diffusion limit (FDL) an if f $\tau_\rho
<<\tau_D$ then we are in the slow diffusion limit(SDL). In these two
limiting cases the eigenvalues and the analytical expressions for the
magnetic signal simplify
considerably. In the FDL the spins traverse the triangular domain many
times before they relax and the magnetic decay is dominated by one
mode. By simply replacing $\tan$ with its argument in equation
(\ref{tranc}), we regain the famous result by Brownstein and
Tarr\cite{brownstein2}, by direct calculation:
\bea
\frac{1}{T_{00}}=\rho\frac{2}{R}=\rho\frac{S}{V}\, .
\eea
We write the magnetic signal from the equilateral triangle as
$M(t)=\sum_i^{\infty}I_{ii}\exp(-t/T_{ii})$, the coefficients and
eigenvalues are summarized in table \ref{tabt2} and \ref{tabi}.
As seen from table \ref{tabi}, the FDL is dominated by one mode and hence a
single decay time. In Figure \ref{fig:Tn} we have plotted the
eigenvalues as a function of $\gamma$.
\begin{table}[t]
\begin{center}
\caption{Characteristic decay times as a function of surface
  relaxtivity  ,  pore radius $R$ and diffusion coefficient $D$ in an
  equilateral  triangle. Note that the lowest mode is independent of
  the diffusion constant  in FDL, while the higher modes are
  independent  of the surface relaxtivity }
\label{tabt2}
\begin{tabular}{|c| c| c|}\hline\hline
          &   Fast Diffusion Limit &  Slow Diffusion Limit   \\
          &   $\tau_D>>\tau_\rho\,(\rho R/D << 1)$& $\tau_D<<\tau_\rho\,(\rho R/D>>1)$ \\ \hline
$(T_2)_{00}$ & $\frac{R}{2\rho}$ & $\frac{9 R^2}{4D\pi^2}$ \\ \hline
$(T_2)_{ii}$ & $\frac{9R^2}{4D\pi^2i^2}$ & $\frac{9 R^2}{4D\pi^2(1+i)^2}$ \\ \hline\hline 
\end{tabular}
\end{center}
\end{table} 

\begin{table}[t]
\begin{center}
\caption{The intensities in the fast and slow diffusion limit. Note that the sum
  of the intensities add up to one as they should, e.g. $\sum_{i=0}^\infty 1/(1+i)^2=\pi^2/6$
  and $\sum_{i=0}^\infty\sin[2(1+i)\pi/3]/(1+i)=\pi/6$. In FDL ($\tau_\rho>>\tau_D$)
  $I_{ii}=\delta_{i,0}$ for all initial spin densities.}
\label{tabi}
\begin{tabular}{|l|c|}\hline\hline
                 &  Slow Diffusion Limit   \\
                 & $\tau_D>>\tau_\rho\,(\rho R/D>>1)$ \\ \hline
$I_{ii}$ uniform &  $\frac{6}{(1+i)^2\pi^2}$ \\ \hline
$I_{ii}$ center  &  $\frac{6\sin[2(1+i)\pi/3]}{(1+i)\pi}$ \\ \hline
$I_{ii}$ side    &  $\frac{4\sin[(1+i)\pi/8]+2\sin[7(1+i)\pi/4]}{(1+i)\pi}$ \\ \hline
$I_{ii}$ corner  &  $\frac{4\sin[7(1+i)\pi/8]+2\sin[(1+i)\pi/4]}{(1+i)\pi}$ \\ \hline\hline

\end{tabular}
\end{center}
\end{table}

\begin{figure*}[t]
\begin{center}
$\begin{array}{c c}
\epsfxsize=7.2cm
\epsffile{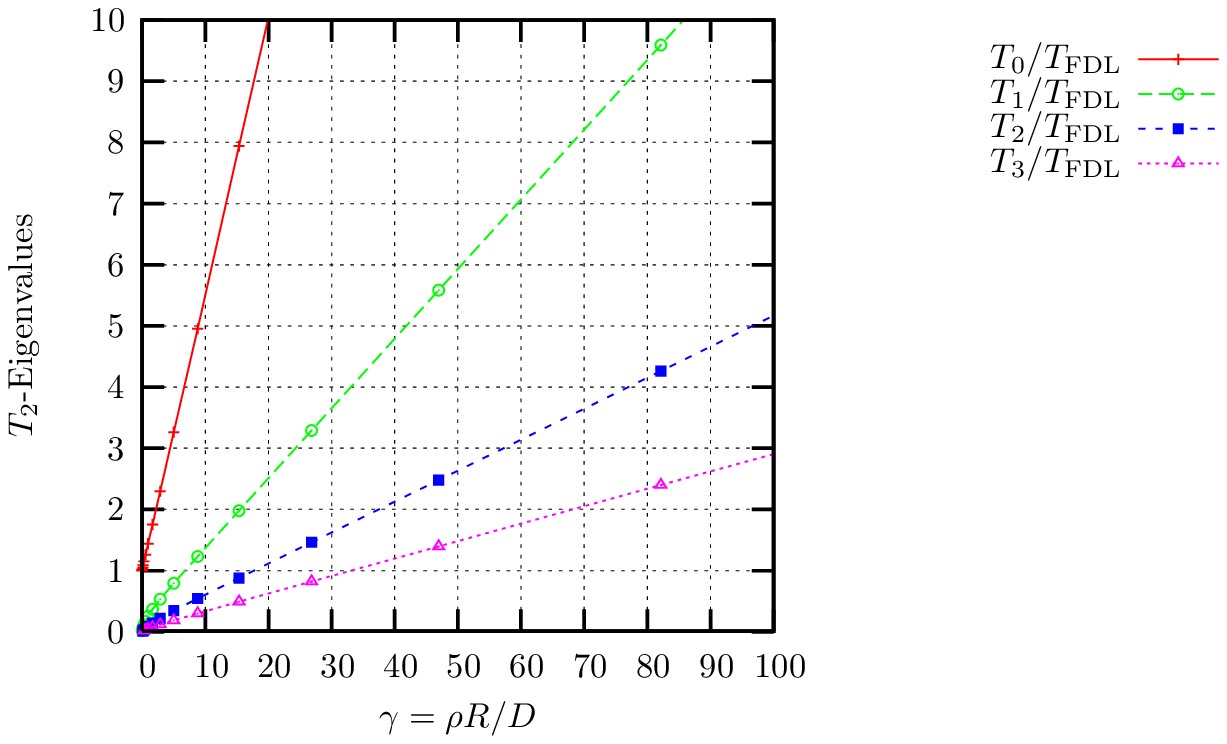} &
    \epsfxsize=7.2cm
    \epsffile{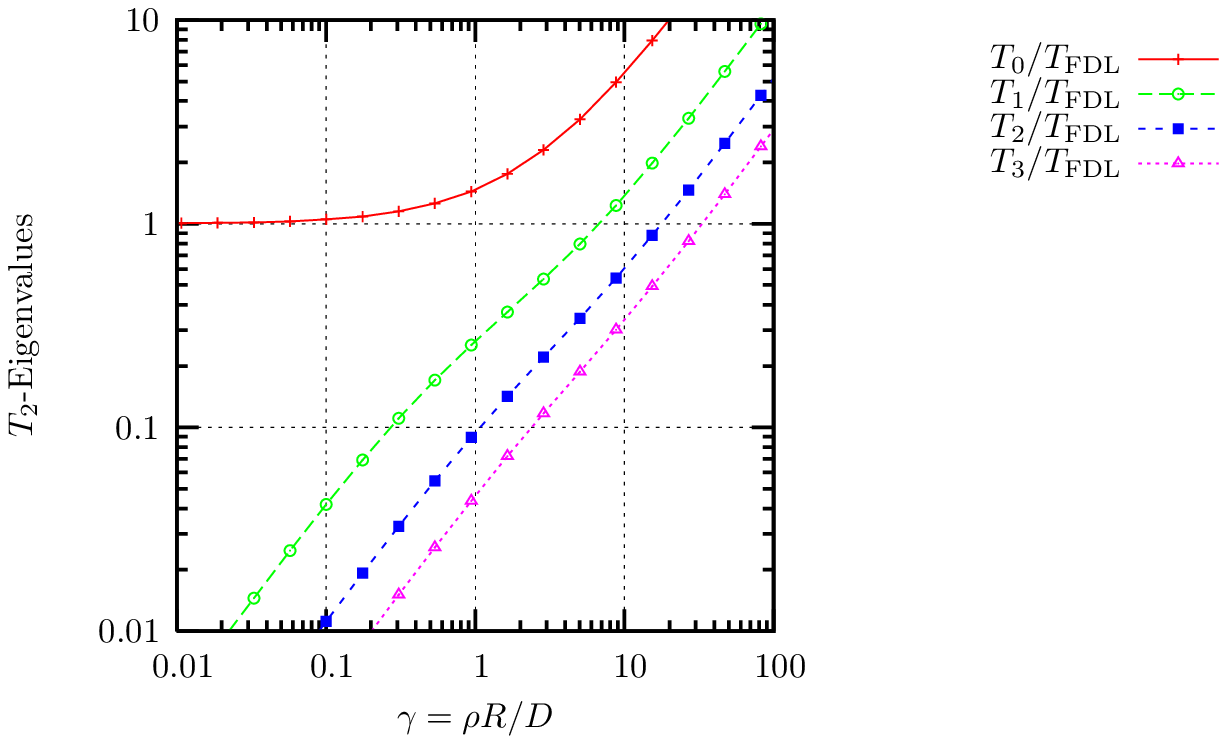}
\end{array}$
\end{center}
\caption{Left: The eigenvalues as a function of $\gamma$,
  $T_{\text{FDL}}=r/(2\rho)$. Right: same as left but log-log plot.
  For $\gamma=0.1$ there is a 10\% deviation and at
  $\gamma=0.4$ there is a 20\% deviation from the FDL.}
\label{fig:Tn}
\end{figure*}

Figure \ref{fig:Tn} and \ref{fig:In} show that the eigenvalues 
deviate faster from the FDL compared to the intensities created with
an uniform initial spin density. This means that the magnetic decay 
is still mono exponential, but the simple relation $T_2^{-1}=\rho S/V$ is violated. 

\section{Numerical Algorithm - Random Walk}\label{sec:rw}

\begin{figure}[t]
\begin{center}
   \epsfig{file=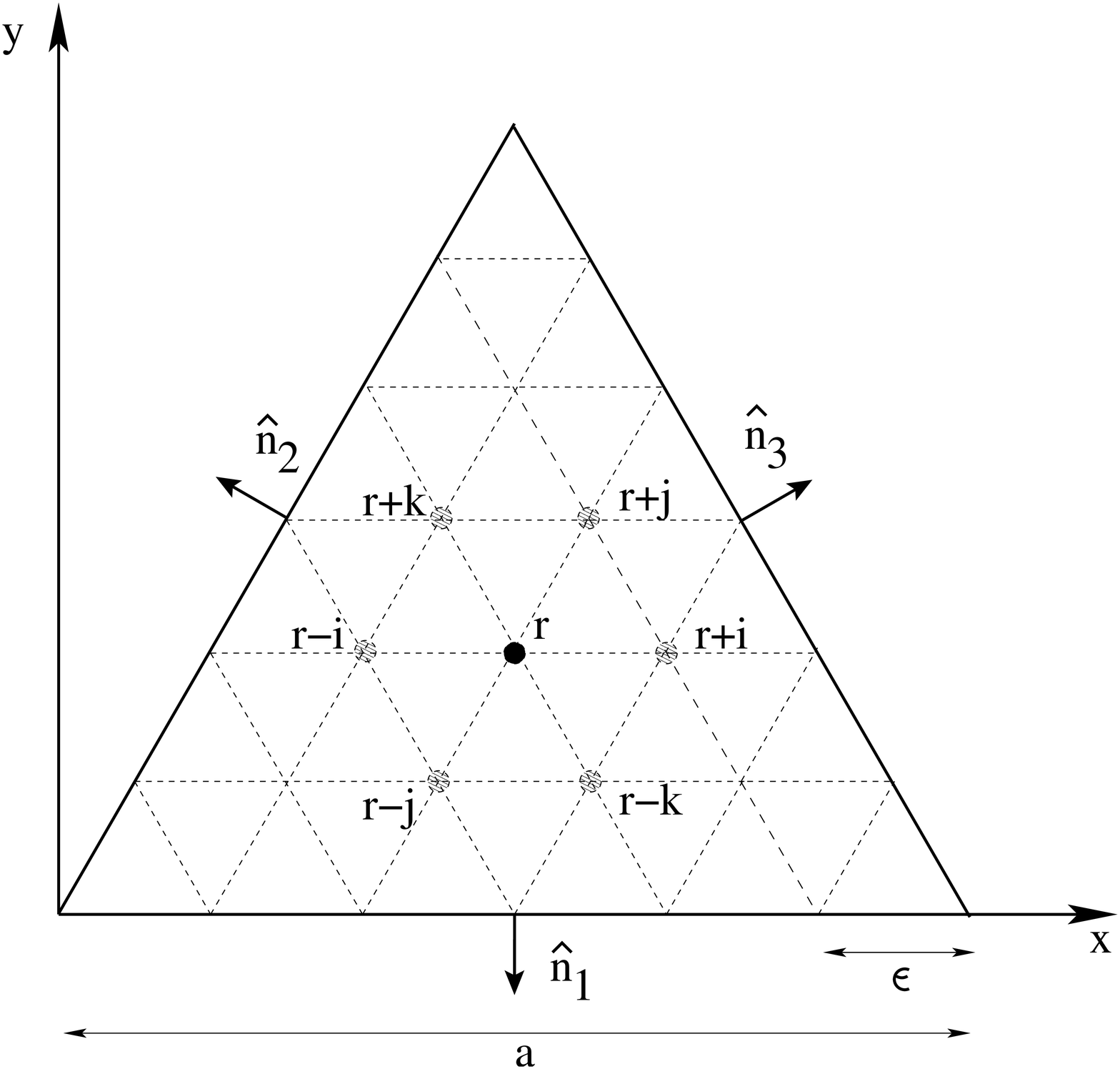,width=5.5cm}
\caption{Grid for random walk simulation, lattice spacing $\epsilon$
  and side length $a$, unit outward normal vectors ${\hat{{\bf n}}}_{1,2,3}$, }
\label{fig:rand}
\end{center}
\end{figure}
In this section we will show explicitly that a random walk algorithm\cite{rwk}
gives rise to the diffusion equation (\ref{diffm}), with the
appropriate boundary condition given in equation (\ref{mbound}), when
the number of grid points approaches infinity. The random walk algorithm
presented here will serve two purposes: (1) it will give a
confirmation on the analytical results and (2) it can be extended to
account for any number of phases inside a triangle. More details on
scaling and convergence in random walk simulations are given in
appendix \ref{a:rw}.  

For random walk simulation in an equilateral triangle we choose 
a hexagonal grid, see Figure \ref{fig:rand}. Random walkers are 
placed at random at a lattice point
in the equilateral triangle. The number of random walkers at an
interior point $r$ ( see Figure \ref{fig:rand}) when the clock advance
one step $\tau$ is then given by :
\bea
&&M(\rvek,t+\tau)-M(\rvek,t)=\no\\&&
 \sxth\left[M(\rvek+\epsilon\ihat,t)-M(\rvek,t)
           +M(\rvek-\epsilon\ihat,t)-M(\rvek,t)\right.\nonumber\\
& & +\left.M(\rvek+\epsilon\jhat,t)-M(\rvek,t)
           +M(\rvek-\epsilon\jhat,t)-M(\rvek,t)\right.\nonumber\\
& & +\left.M(\rvek+\epsilon\khat,t)-M(\rvek,t)
           +M(\rvek-\epsilon\khat,t)-M(\rvek,t)\right]\nonumber\\
& &-\kappa M(\rvek,t)
\eea
The probability for a random walker to take a step in each direction
is 1/6, the probability for a random walker to die during the time step
$\tau$ is $\kappa$. Dividing by $\tau$, we find:
\begin{eqnarray}
&&\difrac{M(\rvek,t+\tau)-M(\rvek,t)}{\tau}=\no\\&&
 \difrac{\epsilon^2}{6\tau}\left[\difrac{M(\rvek+\epsilon\ihat,t)-2M(\rvek,t)
           +M(\rvek-\epsilon\ihat,t)}{\epsilon^2}\right.\nonumber\\
&&+\left.\difrac{M(\rvek+\epsilon\jhat,t)-2M(\rvek,t)
           +M(\rvek-\epsilon\jhat,t)}{\epsilon^2}\right.\nonumber\\
&&+\left.\difrac{M(\rvek+\epsilon\khat,t)-2M(\rvek,t)
           +M(\rvek-\epsilon\khat,t)}{\epsilon^2}\right]\nonumber\\
& &-\difrac{\kappa}{\tau} M(\rvek,t)
\end{eqnarray}
The lattice spacing is given by $\epsilon$. Taking the limit
$\tau\to 0$, $\epsilon\to 0$, $\kappa\to 0$:
\begin{equation}
\didrv{M}{t}=\difrac{\epsilon^2}{6\tau}\left[
             \didrv{^2M}{i^2}+\didrv{^2M}{j^2}+\didrv{^2M}{k^2}\right]
            -\difrac{\kappa}{\tau} M(\rvek,t)\, .
\end{equation}
Changing from lattice coordinates to Cartesian coordinates:
\begin{equation}
\didrv{^2}{i^2}+\didrv{^2}{j^2}+\didrv{^2}{k^2}
  =\difrac{3}{2}\left(\didrv{^2}{x^2}+\didrv{^2}{y^2}\right)\, ,
\end{equation}
we finally arrive at :
\bea
&&\didrv{M}{t}=D\lplc M-\frac{M}{{T_2}_b}, \quad\text{where}\label{diffrw}\\
&&D=\difrac{\epsilon^2}{4\tau}\qquad\text{and}\qquad
\frac{1}{{T_2}_b} =\difrac{\kappa}{\tau}\,.
\label{taud}
\eea
This is the same equation as (\ref{diffm}). At a lattice point at a
boundary surface normal to the ${\hat{\bf{n}}}_3$-vector (in Figure
\ref{fig:rand}), the equation becomes:
\begin{eqnarray}
&&\epsilon\difrac{M(\rvek,t+\tau)-M(\rvek,t)}{\tau}=\no\\
&&\difrac{2}{3}\epsilon
 \difrac{\epsilon^2}{4\tau}\left[\difrac{M(\rvek+\epsilon\khat,t)-2M(\rvek,t)
           +M(\rvek-\epsilon\khat,t)}{\epsilon^2}\right]\nonumber\\
& & +\difrac{2}{3}\difrac{\epsilon^2}{4\tau}
     \difrac{M(\rvek-\epsilon\ihat,t)-M(\rvek,t)}{\epsilon}\nonumber\\
& & +\difrac{2}{3}\difrac{\epsilon^2}{4\tau}
     \difrac{M(\rvek-\epsilon\jhat,t)-M(\rvek,t)}{\epsilon}\nonumber\\
& & -\difrac{1}{3}\difrac{\zeta\epsilon}{\tau}M(\rvek,t)\nonumber\\
& & -\epsilon\difrac{\kappa}{\tau}M(\rvek,t)\, .\label{likn}
\end{eqnarray}
We have assumed that the walkers have a probability $\zeta$ of being
killed when stepping from an interior  point to the wall. If a walker
is not killed it is assumed
to return to the interior point in the same time step.
In the limit $\tau\rightarrow0$, $\epsilon\rightarrow0$, the fractions
involving $M$ are recognized to approach $\partial M/\partial t$,
$\partial^2M/\partial\epsilon^2$, $-\partial M/\partial i$ and
$-\partial M/\partial j$, respectively.
Using the following relation:
\begin{equation}
\didrv{}{i}+\didrv{}{j}=\sqrt{3}\;{\bf \hat{n}}_3\cdot\nabla
\end{equation}
and as the LHS  and the first term on the RHS in equation (\ref{likn}) are of higher
order in $\epsilon$, they can be neglected compared to the
others, hence :
\be
D\bs{n}_3 \cdot\nabla M +\rho M=0\qquad\text{and}\qquad
\rho=\difrac{1}{2\sqrt{3}}\difrac{\zeta\epsilon}{\tau}\, .
\label{boundrw}
\ee
By symmetry the other boundary sides give the same answer and we then
regain equation (\ref{mbound}).

Using the fact that number of lattice points
from corner to corner along an edge is
\be
N=1+a/\epsilon\, ,
\label{epsilon}
\ee
from equation \eqn{taud} and \eqn{boundrw} we find  the following important relation:
\begin{equation}
\gamma=\frac{\rho R}{D}=\difrac{1}{3}(N-1)\zeta
\label{gzrel}
\end{equation}
For a given $\gamma$ the probability for a random walker to
die when it hits the wall can be calculated from
equation(\ref{gzrel}).
Consider the ratio of bulk lattice points to
surface lattice points we can define a new parameter $\theta$, which
holds the information of bulk relaxation $T_{2b}$:
\begin{eqnarray}
\theta&=&\frac{N}{2}\frac{1+1/N}{(1-1/N)^2}\frac{\kappa}{\zeta}
=\frac{1}{4\sqrt{3}}\frac{a}{\rho {T_2}_b}\frac{N^2(N+1)}{(N-1)^3}
        \nonumber\\
    &\begin{CD} @>>{\epsilon\to 0}>\end{CD}&
       \frac{1}{4\sqrt{3}}\frac{a}{\rho {T_2}_b}
      =\sxth\dfrac{{\di}\frac{a^2}{4D}}{\gamma {T_2}_b}\, .\label{theta1}
\end{eqnarray}
The magnetic signal can then be calculated numerically by placing  a
number of random walkers inside the equilateral
triangular domain. For uniform initially spin density the walkers are
placed at random and for the delta densities all the walkers start in
a point of the triangle. At each time step the walkers take one step of
length $\epsilon$ and dies with a probability $\kappa$. If the walkers
hit the wall they die with probability $\zeta$. The magnetic signal
will then be proportional to the number of walkers alive at each time step
$\tau$. The lattice spacing and time is related to the physical length
and time by using equations (\ref{taud}).

\begin{figure*}[t]
\begin{center}
$\begin{array}{c c}
\epsfxsize=7.2cm
\epsffile{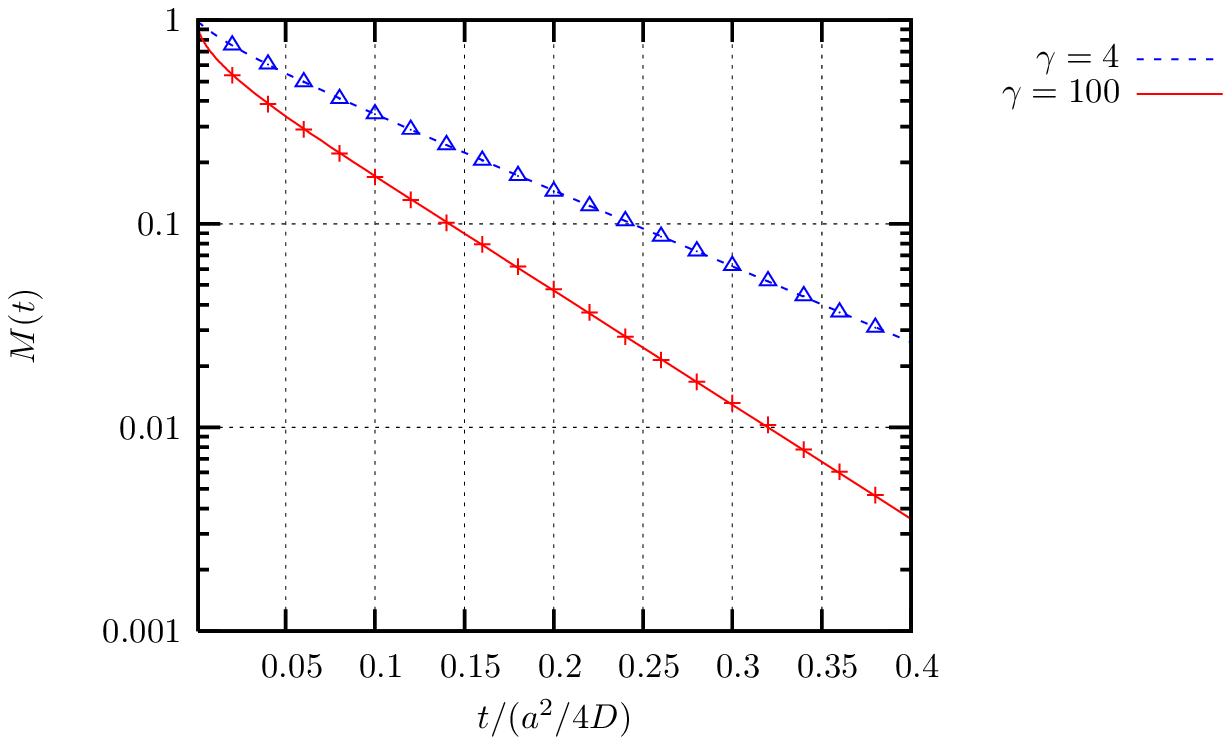} &
    \epsfxsize=7.2cm
    \epsffile{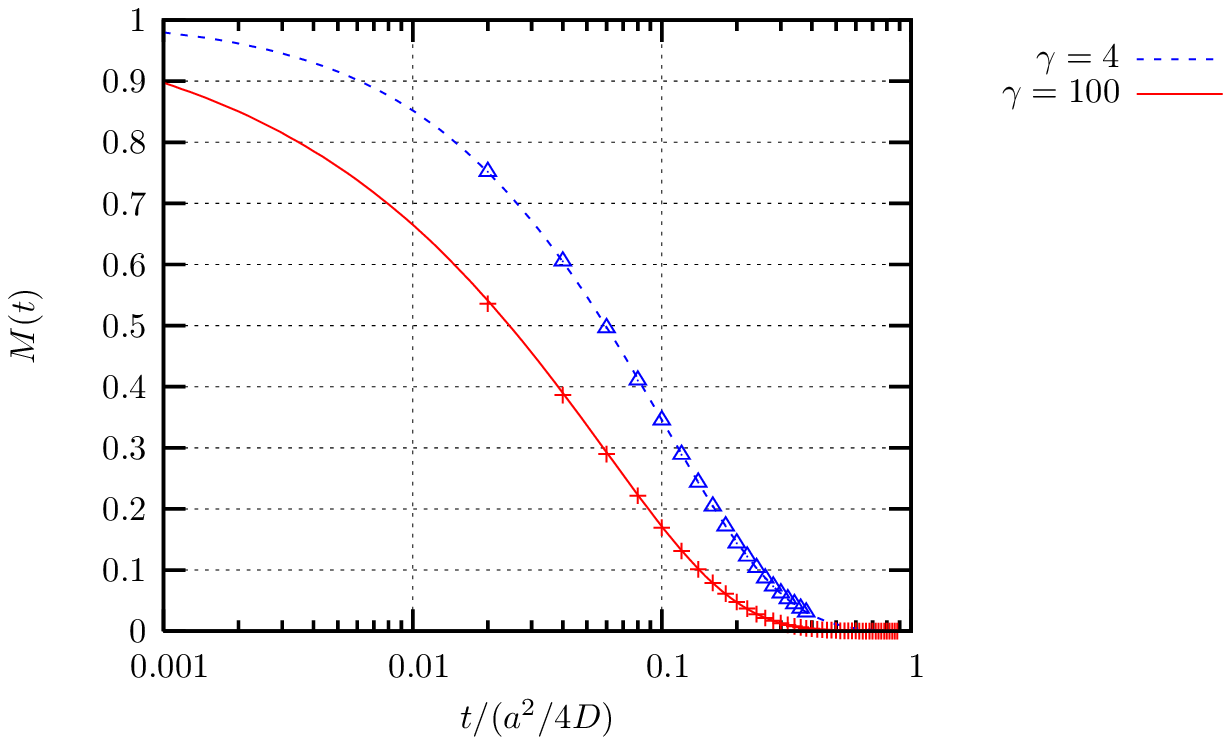}
\end{array}$
\end{center}
\caption{Left:Comparison of random walk simulation (points) and
  analytical solution (line) for $\gamma=4$ and 100. $N=501$ and $10^6$ random walkers. The
  initial spin density is uniform. Right:
  Same as left but log scale on the x-axis instead of the y-axis.}
\label{fig:rw}
\end{figure*}
\begin{figure*}[t]
\begin{center}
$\begin{array}{c c}
\epsfxsize=7.2cm
\epsffile{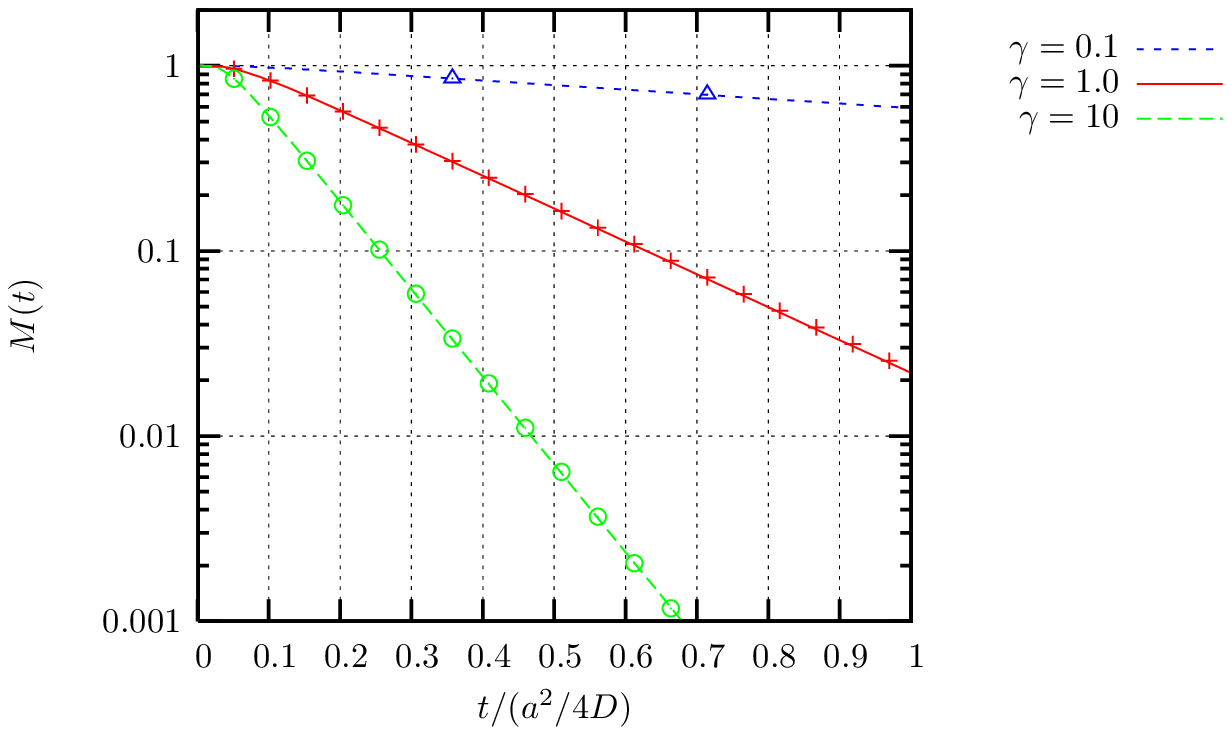} &
    \epsfxsize=7.2cm
    \epsffile{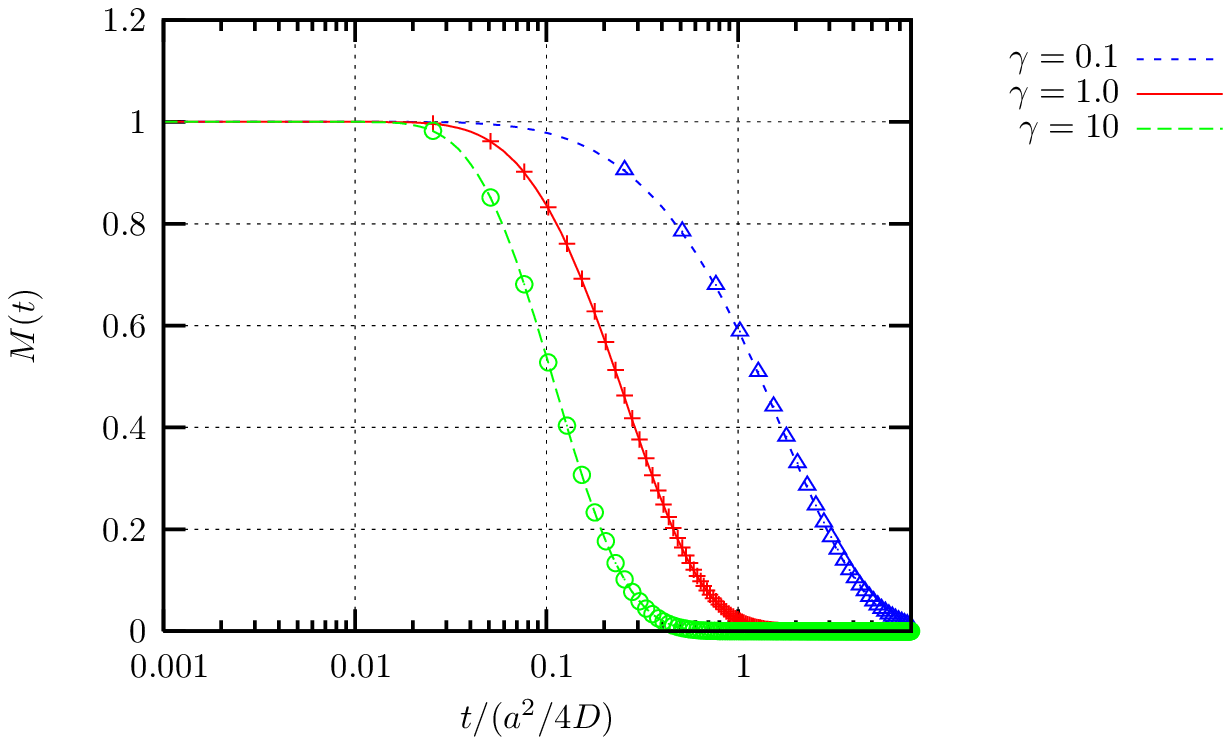}
\end{array}$
\end{center}
\caption{Left:Comparison of random walk simulation (points) and
  analytical solution (line) for $\gamma=0.1, 1$ and 10. The
  initial spin density is the center density. $N=501$ and $10^6$ random walkers. Right:
  Same as left but log scale on the x-axis instead of the y-axis.}
\label{fig:crw}
\end{figure*}
In Figure \ref{fig:rw} and \ref{fig:crw} we have compared numerical
and analytical results. There is clearly a very good match. The
largest deviation between numerical and analytical results are for
long times. This is natural as the number of walkers is low and the
statistic is poorer. Note that
if one extrapolate the straight line for high $\gamma$ values in
Figure \ref{fig:rw} (left) and \ref{fig:crw} (left), it crosses the
y-axis at $6/\pi^2$ and $6/\pi$, which is expected from Table \ref{tabi}.
 We have also made some comparison with $\theta\ne
0$, i.e. with bulk relaxation (see equation (\ref{theta1})).
In Figure \ref{fig:theta}, we have plotted results for $\theta=0.1$ and 1.
For $a=100\mu m$, $D=2500 \mu m^2/s$ and $\gamma=1$,  $\theta=0.1, 1$
corresponds to a bulk relaxation of 1.7 and 0.17~$s$ respectively.

\begin{figure*}[t]
\begin{center}
$\begin{array}{c c}
\epsfxsize=7.2cm
\epsffile{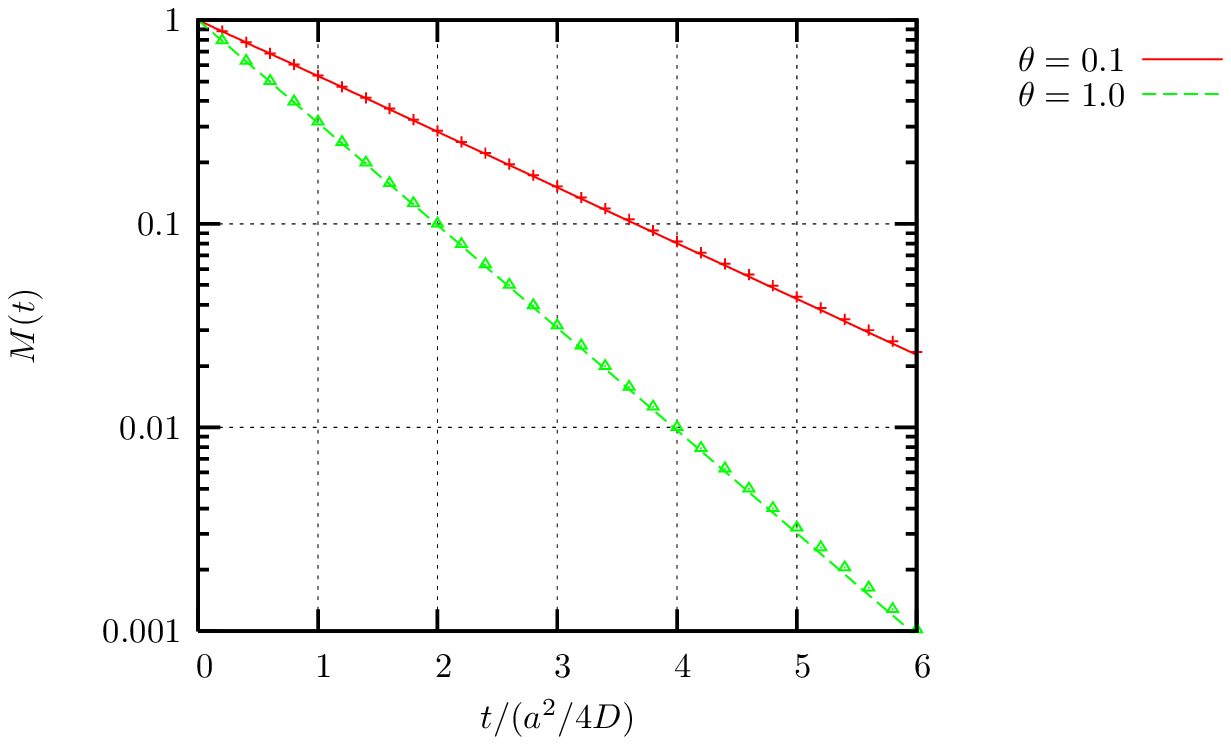} &
    \epsfxsize=7.2cm
    \epsffile{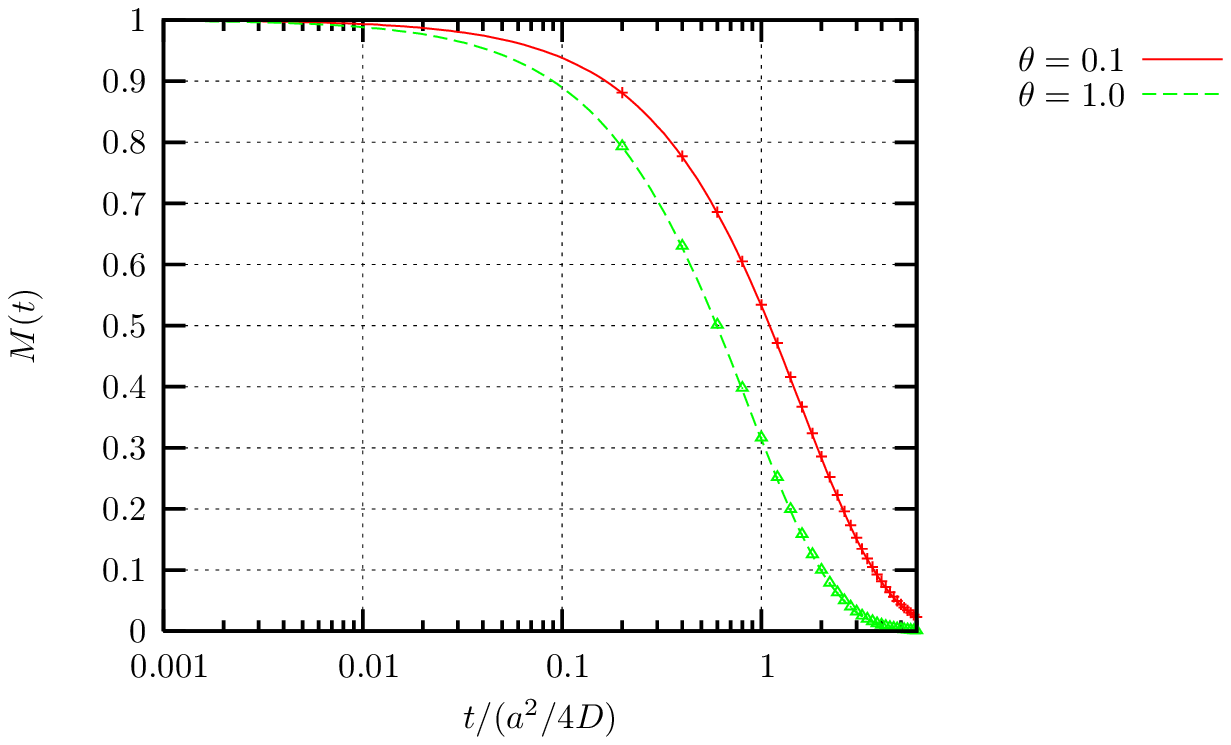}
\end{array}$
\end{center}
\caption{Left:Comparison of random walk simulation (points) and
  analytical solution (line) for $\theta=0.1, 1$, $\gamma=1$. Uniform
  initially spin density,  $N=201$ and $10^6$ random walkers. Right:
  Same as left but log scale on the x-axis instead of the y-axis.}
\label{fig:theta}
\end{figure*}

\subsection{Scaling in random walk simulations}
Eliminating $\epsilon$ between equation (\ref{taud})
and equation (\ref{epsilon}) gives a
relation between the intrinsic time scale for diffusion and the time step
length in a simulation with a given $N$ value:
\begin{equation}
\tau_D\equiv\tau(N-1)^2=\difrac{a^2}{4D}\label{tau_d}
\end{equation}
If, for given $\gamma$ and $\theta$ values, the average magnetization is
plotted not against the actual time $t$ or the integer time step counter
$t/\tau$ but in units of the intrinsic time scale, viz., a plot against
\begin{displaymath}
\difrac{t}{\tau_D}=
\difrac{t}{\tau(N-1)^2}
\end{displaymath}
then the results for various $\{N, \zeta\}$ consistent with the given $\gamma$
value should be expected to plot on top of each other, provided inaccuracies
due to finite $N$ values are unimportant. This follows since
$t/\tau\propto(N-1)^2$, where $N$ is not a physical parameter but an
artificial one determined by the simulation conditions.  
Technically, there would thus be a scaling
property in the limit $N\rightarrow\infty$, in that the magnetization should
depend only on a certain algebraic combination of simulation parameters.

The predicted scaling has been checked numerically for several $\gamma$ values,
using $\theta=0$ (no bulk relaxation), and has indeed been found to hold.
(The origins and magnitude of deviations from exact scaling is
discussed in more detail in Appendix \ref{a:rw}.)
Its practical value is that for a given simulation, one may choose that set of
values for $\{N, \zeta\}$ which gives an optimally acceptable combination
of accuracy and computing time requirements. This scaling is also very
useful when comparing numerical and analytical results, once the
results have been normalized by $t\to t/\tau_D$ only one parameter,
$\gamma$, has to be given. Pore size $a$, diffusion constant $D$ and
surface relaxtivity $\rho$ are unimportant as long as the combination
$\gamma=\rho R /D = \rho a/(2\sqrt{3} D)$ stay unchanged.  

This scaling is also consistent with the one found by Valfouskaya et
al.\cite{vf1,vf2,vf3}. They show that for zero surface relaxtivity an
universal curve exists for a porous media. It is universal in the
sense that different saturations give the same curve. 
Considering magnetic signal from a single pore, the curve by Valfouskaya et
al. reduce to the effective
diffusion constant normalized by the molecular diffusion constant
($D$), $D_{\text{eff}}(T_l)/D\equiv \langle \bs{r}(T_l)^2\rangle /(4t
D)$, where $T_l=4/(9\sqrt{\pi})\sqrt{D t}/l_s$ . $l_s$ is a typical
  length scale of the porous media, in our case it is
  $l_s=V/S=a/(4\sqrt{3})$ and $T_l=2/(9\sqrt{\pi})\sqrt{t/\tau_D}$,
  where $\tau_D$ is defined in equation (\ref{tau_d}).   
We have showed, in the beginning of this subsection,  that once $\gamma$ is fixed (and
zero bulk relaxtivity) all magnetization curves, independent of the
initial spin density, will plot on top of each other when the physical
time has been rescaled according to  $t\to t/\tau_D$. This is exactly the type of scaling
used by \cite{vf1,vf2,vf3}. 

\section{Discussion}

In pore network modeling literature, the introduction of sharp
corners has shown to be very fruitful in order to understand
multi phase behavior. In this paper we have considered magnetic signal
from one of the simplest geometries containing sharp corners, the
equilateral triangle. This solution is similar to the one dimensional
geometries: plate, cylinder and sphere in the case of uniform initial
magnetization. When the initial magnetization is non uniform the
triangle behaves qualitatively different. In \cite{song} it was shown
experimentally how to generate a non uniform initial magnetization,
due to diffusion in internal field (DDIF). A reference signal with
uniform initial magnetization was used and a signal due to a non uniform
initial magnetization was created. It was seen that the non uniform
magnetization was highly multi exponential. With the triangular
geometry this result can be explained by calculating the magnetic
signal with uniform initial magnetization and magnetization
concentrated close to the corners, shown in Figure \ref{fig:nu}.
\begin{figure}[t]
\begin{center}
   \epsfig{file=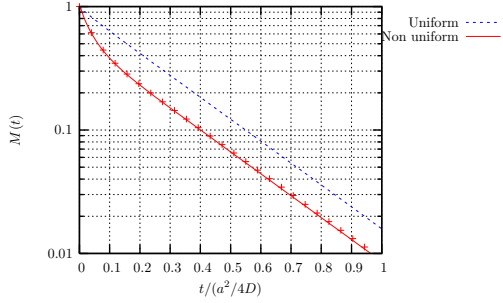,width=7cm}
\caption{Magnetic signal for uniform and corner initial
  spin density ($\gamma=1$). The points are random walk calculations and lines
  analytical.}
\label{fig:nu}
\end{center}
\end{figure}
Compared with plate, cylinder and sphere only the triangle with non
uniform initial magnetization close to the corners had a magnetic
signal which behaved multi exponential for $\gamma=1$.

In order to fully explain the experimental results generated by the
DDIF technique, effect of pore connectivity\cite{zielenski} and internal gradients
must also be considered. However, the results shown in Figure \ref{fig:nu}
are very striking and they seem to capture some of the essence in the
experimental results\cite{song}. 

The numerical results presented here can be extended to account for
more than one phase inside the triangle. These results will then be
used to study the influence of wettability on the magnetic signal and
will form a basis when constructing inversion
integrals used to interpret NMR-logs.

\section{Conclusion}

We have presented an analytical solution for the magnetic signal from
an equilateral triangular pore, with surface relaxation.
To our knowledge this solution has not
been presented before. This solution will be used in theoretical
studies of single- and two-phase NMR response from equilateral
triangles, which can be used as basic building blocks for pore scale
models. This solution is also very important when interpreting
experiments done at equilateral triangular tubes, which is in progress
at our group.

We have studied how the magnetic signal changes for different
initial conditions. For non uniform initial magnetization the 
equilateral triangular geometry behaves qualitatively different 
than plate, cylinder and sphere due to the sharp corners.  

In a forthcoming paper we will also present results for the pulsed
field gradient spin echo (PFGSE)
sequence\cite{cpgm,cal1,price1,price2}, the single-phase result can be calculated from
the Green's function presented in this paper. Two-phase results for any
value of $\gamma$ must in general be solved by a random walk algorithm, except
for the limit $\gamma\to 0$ where analytical results can be found. The
analytical result will then serve to calibrate the numerical
random walk algorithms for two-phase.

\vspace{1 cm}
{\bf Acknowledgments}\\
The authors acknowledge ConocoPhillips and the Ekofisk Coventurers,
including TOTAL, ENI, Hydro, Statoil and Petoro, for financing the work and
for the permission to publish this paper from the research center
COREC.

\appendix

 \section{The Full Green's Function for an Equilateral Triangle}
\label{appa}
The Green's function for the equilateral triangle is given by the
following equation.
\bea
&&\Gx=\sum_{i=0}^{\infty}\phi^s_{i,i}\ax\phi^s_{i,i}\axp e^{-t/T_{i,i}}
\no\\
&&+\sum_{i=0}^{\infty}\sum_{j=i+1}^{\infty}\left[\phi^s_{i,j}\ax\phi^s_{i,j}\axp
+\phi^a_{i,j}\ax\phi^a_{i,j}\axp\right]e^{-t/T_{i,j}}\\
&&\phi^{s,a}_{i,j}\equiv\fr{T^{s,a}_{i,j}}{N^{s,a}_{i,j}}\quad\text{and}\quad
N^{s,a}_{i,j}\equiv\int_\Delta dxdy\; T^{s,a}_{i,j}(x,y)\,T^{s,a}_{i,j}(x,y)\, .
\eea
The complete set of orthogonal eigenfunctions for the equilateral triangular
triangle are\cite{mccartin3}:
\bea
&&T_{ij}^s(x,y)=\cos\left[\fr{\pi\lambda}{3R}(3R-y)-\delta_1\right]\,
\cos\left[\fr{\sqrt{3}\pi(\mu-\nu)}{9R}(x-\sqrt{3} R)\right]
\\&&+\cos\left[\fr{\pi\mu}{3R}(3R-y)-\delta_2\right]\,
\cos\left[\fr{\sqrt{3}\pi(\nu-\lambda)}{9R}(x-\sqrt{3} R)\right]
+\cos\left[\fr{\pi\nu}{3R}(3R-y)-\delta_3\right]\,
\cos\left[\fr{\sqrt{3}\pi(\lambda-\mu)}{9R}(x-\sqrt{3} R)\right]\no\\
&&T_{ij}^a(x,y)= 
\cos\left[\fr{\pi\lambda}{3R}(3R-y)-\delta_1\right]\,
\sin\left[\fr{\sqrt{3}\pi(\mu-\nu)}{9R}(x-\sqrt{3} R)\right] \\
&&+\cos\left[\fr{\pi\mu}{3R}(3R-y)-\delta_2\right]\,
\sin\left[\fr{\sqrt{3}\pi(\nu-\lambda)}{9R}(x-\sqrt{3} R)\right]
+\cos\left[\fr{\pi\nu}{3R}(3R-y)-\delta_3\right]\,
\sin\left[\fr{\sqrt{3}\pi(\lambda-\mu)}{9R}(x-\sqrt{3} R)\right]\no\, .
\eea
$R=a/(2\sqrt{3})$ is the radius of the inscribed circle in an
equilateral triangle of side $a$.  The index $s,a$ of the
eigenfunctions refers to symmetric,antisymmetric about the line $x=a/2$,
respectively. The complete set of orthogonal eigenfunctions is 
$\{T_{ij}^s(i\ge j) ;\, T_{ij}^a(i>j)\}$, for further details on this
point see\cite{mccartin3}. The constants
$\mu,\nu,\lambda,\delta_1,\delta_2,\delta_3$ are determined by solving
the following three transcendental equations originating from the
boundary condition (\ref{phi}):
\bea
&&\left[2L-M-N-(i+j)\pi\right]\tan L\;=\;3\gamma ,\, -\fr{\pi}{2}<L\le 0\no\\
&&\left[2M-N-L+i\pi\right]\tan M\;=\;3\gamma ,\, 0<M\le\fr{\pi}{2}\no\\
&&\left[2N-L-M+j\pi\right]\tan N\;=\;3\gamma ,\, 0<N\le\fr{\pi}{2}\, ,
\label{bound}
\eea
where $\gamma=R\rho/D$, $i=0,1,\ldots$, $j=i,i+1,\ldots$ and the
auxiliary variables $L,\,M$ and $N$ are related to
$\mu\,,\nu\,,\delta_{1,2,3}$ in the following way: 
\bea
&&\delta_1=L-M-N\;\;,\; \delta_2=-L+M-N\;\;,\;\delta_3=-L-M+N\no \\
&&\mu=\fr{2M-N-L}{\pi}+i\;\;,\; \nu=\fr{2N-L-M}{\pi}+j\;\;,\;\lambda=-\mu-\nu\;.
\label{dmurel}
\eea
Finally, the eigenvalues are given by:
\be
T_{ij}^{-1}=\fr{2D}{27}\left(\fr{\pi}{R}\right)^2\left[\lambda^2+\mu^2+\nu^2\right]\, .
\ee
\bea
&&{N^{s,a}}^2=\int_0^{3R}\int_{y/\sqrt{3}}^{-y/\sqrt{3}+2\sqrt{3}R} dxdy
T_{s,a}(x,y)T_{s,a}(x,y)
\eea
in an obvious notation, the subscript $i,j$ has been suppressed. After
a rather lengthy manipulation, the result can be written:
\bea
&&{N_{ij}^\prime}^2\equiv {N^s}^2={N^a}^2=F[\mu, \nu, \delta_1, \delta_2]
+ F[\nu, \mu, \delta_1, \delta_3]+  F[\nu, -\mu - \nu, \delta_2, \delta_3]\no\\&&
+ Q[\mu, \delta_2] + Q[\nu, \delta_3]+ Q[-\mu - \nu, \delta_1]
\,\quad i\neq j\, ,
\eea
and for $i=j$:
\bea
&&{N^s_{ii}}^2= -\frac{9 {\sqrt{3}} {R^2}}{16 {\mu^2} {{\pi }^2}}
 \big\{-8-8 {\mu^2} {{\pi }^2}+7 \cos [2 {\delta_2}]+8
\cos [2 \mu \pi ]\no\\&&\qquad+\cos [2 {\delta_2}-4 \mu \pi ]
-8 \cos [2 {\delta_2}-2 \mu \pi
]-  4 \mu \pi  \sin [2 {\delta_2}]+16 \mu \pi  \sin [2 {\delta_2}-2\mu
  \pi ]\big\}\label{anii}
\eea
The functions $F$ and $Q$ are given below:
\bea
&F[\mu, \nu, \delta_1, \delta_2] &=
-\fr{3\sqrt{3}R^2}{4\mu\nu^2(\mu +  \nu)\pi^2}\left\{
\nu\cos[\delta_1 - \delta_2](\nu+\mu \cos [2 (\mu+\nu) \pi])\no\right.\\
&&\left.+(\mu+\nu) (-\nu \cos [\delta_1-\delta_2+2 \mu \pi
]
-\mu\cos[{\delta_1} +{\delta_2}]\no\right.\\&&\left.+\mu\cos
[{\delta_1}+{\delta_2}+2 \nu \pi ]+2 \mu\nu \pi  \sin
[{\delta_1}+{\delta_2}])
- \sin [{\delta_1}-{\delta_2}] \sin [2 (\mu+\nu) \pi ]\right\}\label{fa}\\
&Q[{\mu},{\delta_2}]&=\frac{3 {\sqrt{3}} {R^2}}{8 {\mu^2} {{\pi }^2}} (\cos [2
({\delta_2}-\mu \pi )]-\cos [2 {\delta_2}]+2 \mu \pi  (\mu \pi -\sin [2 ({\delta_2}-\mu \pi )]))
\label{qa}
\eea
Note that when $i=j$ we have $M=N,\, \delta_2=\delta_3,\,\mu=\nu$ and
$2\pi\mu=\delta_2-\delta_1+2i\pi$. The eigenfunctions and
transcendental equations simplifies and can after some manipulations be
reduced the equations given in section \ref{sec:green}.
 
To summarize the Green's function for the general case can be written :
\bea
&&\Gx=\sum_{i=0}^{\infty}\frac{T^s_{i,i}\ax T^s_{i,i}\axp}{N_{ii}^2} e^{-t/T_{i,i}}
\no\\
&&+\sum_{i=0}^{\infty}\sum_{j=i+1}^{\infty}\frac{1}{{N_{ij}^\prime}^2}
\left[T^s_{i,j}\ax T^s_{i,j}\axp
+T^a_{i,j}\ax T^a_{i,j}\axp\right]e^{-t/T_{i,j}}
\eea
This function can be used to calculate the magnetic signal when there
are gradients present in the system.

In section \ref{sec:green} we used the fact that only the symmetric
diagonal modes contribute to the magnetic signal. That the
antisymmetric modes does not contribute follows by definition, but the
non diagonal symmetric modes ($i\neq j$) may give a non-vanishing contribution to the
magnetic signal. To show this we 
integrate the symmetric modes over the equilateral triangular domain :
\bea
&&A=\int_0^{3R}\int_{y\sqrt{3}}^{-y\sqrt{3}+2\sqrt{3}R}\,dxdy
T_{s}(x,y)
=-\frac{27\sqrt{3}R^2}{2\pi^2(\mu -\nu)(2 \mu + \nu)( \mu + 2
  \nu)}
\big\{ (\mu - \nu) \cos[\delta_1] + (\mu + 2 \nu) \cos[\delta_2]\no\\&& -
              (2 \mu+\nu) \cos[\delta_3]+
              (\mu+2 \nu) \cos[\delta_3 + \frac{2\pi}{3}( \mu - \nu)] -
              (2 \mu+\nu) \cos[\delta_2 + \frac{2\pi}{3}(-\mu + \nu)] +
              (\mu- \nu) \cos[\delta_2 - \frac{2\pi}{3}(\mu + \nu)]\no\\&& +
              (\mu +2 \nu) \cos[\delta_1 + \frac{2\pi}{3}(2 \mu + \nu)]+
              (\mu- \nu) \cos[\delta_3 - \frac{2\pi}{3}( \mu + 2\nu\pi)]
             -(2 \mu + \nu) \cos[\delta_1 + \frac{2\pi}{3}(\mu + 2 \nu)]\big\}
\eea
From equation (\ref{dmurel}), we find:
\bea
&\mu&= -\frac{\delta_1 - 2\delta_2 + \delta_3}{2\pi} +i\no\\
&\nu&= -\frac{\delta_1 - 2\delta_3 + \delta_2 }{2\pi} - j
\eea
We need to consider two cases separately, $j=i$, $j=i \mod 3 (i\ne
j)$. For $j=i$ we have in addition $\delta_3=\delta_2$ ($\mu=\nu$). We
then find:
\bea
&&A^{i=j}=\frac{9\sqrt{3}R^2}{2\mu^2\pi^2}\{\cos[\delta_2]
- \cos[\delta_2 + 2i\pi- 2\mu \pi] + 2\mu\pi\sin[\delta_2]\}\label{acpgm}
\eea
For  $j=i \mod 3 (j\ne i)$, we write $j=i+3k$, $k=1,2,\ldots$:
\bea
&&A^k=\frac{18\sqrt{3}R^2 }{(\delta_2 - \delta_3 - 2k\pi)(-\delta_1+
    \delta_2 + 2(i + k)\pi)(-\delta_1+ \delta_3 + 2(i + 2k)\pi)}\no\\
&&\times\{(-\delta_2 + \delta_3 +
  2k\pi)\cos[\delta_1] + (\delta_1 -\delta_3 - 2(i +
  2k)\pi)\cos[\delta_2]+ (-\delta_1 + \delta_2 + 2(i +
  k)\pi)\cos[\delta_3]\}
\eea
For the other cases, we find $A=0$. Using the
constraints imposed by the boundary conditions (\ref{bound}) it turns
out that $A^k=0$, this has been verified numerically by
solving equation  (\ref{bound}) for different values of $\gamma=\rho
r/D$. We are then left with equation (\ref{acpgm}) as the final result.

\section{Random Walk}
\label{a:rw}  
\subsection{The relative importance of relaxation types}

Let $\theta$ denote the ratio of bulk relaxation and boundary relaxation rates.
In the fast diffusion limit, with a comparatively flat distribution of
walkers, $\theta$ can be estimated by combining the ratio of bulk lattice
points\footnote{\em The boundary points counted among them here,
since both $\zeta$ and $\kappa$ small in the limit $N\rightarrow\infty$
($\epsilon\rightarrow0$). If not, $(N-2)(N-3)/2$ to be used; anyway no
difference in the limit $\epsilon\rightarrow0$.}
($N(N+1)/2$) to boundary lattice points ($3(N-1)$) with the appropriate ratio
of relaxation probabilities
(a boundary walker will suffer relaxation with probability $\zeta/3$ on this
lattice except in the very corner positions, where the probability is
$2\zeta/3$):
\begin{eqnarray}
\theta&=&\frac{N}{2}\frac{1+1/N}{(1-1/N)^2}\frac{\kappa}{\zeta}
=\frac{1}{4\sqrt{3}}\frac{a}{\rho {T_2}_b}\frac{N^2(N+1)}{(N-1)^3}
        \nonumber\\
    &\begin{CD} @>>{\epsilon\to 0}>\end{CD}&
       \frac{1}{4\sqrt{3}}\frac{a}{\rho {T_2}_b}
      =\sxth\dfrac{{\di}\frac{a^2}{4D}}{\gamma {T_2}_b}\label{theta}
\end{eqnarray}
As follows from inspection of equation (\ref{diffrw}) and
(\ref{boundrw}), and implicitly
stated by Brownstein and Tarr\cite{brownstein1,brownstein2}
the solutions for the average magnetization $M$ will be three-parameter
functions.
In addition to $\gamma$ and $\theta$, one of the intrinsic time scales,
$\tau_D$ for diffusion and $\tau_\rho$ for relaxation at the boundary,
respectively, may be chosen as a parameter:
\bea
&&\tau_D=
\difrac{a^2}{4D}\no\\&&=1.25\:\mbox{s}\times
                 \left(\difrac{a}{0.1\:\mbox{mm}}\right)^2
                 \left(\difrac{2\times10^{-5}\:\mbox{cm}^2\mbox{/s}}{D}\right)
\eea
\bea
&&\tau_\rho=
\difrac{a^2}{4D}/\gamma\no\\&&=2\sqrt{3}\difrac{a}{\rho}
                 =17.32\:\mbox{s}\times
                 \left(\difrac{a}{0.1\:\mbox{mm}}\right)
                 \left(\difrac{5\:\mu\mbox{m}\mbox{/s}}{\rho}\right)
\eea
One such set of parameters may thus be
\begin{displaymath}
\{\gamma, \difrac{a^2}{4D}, \theta\}
\end{displaymath}
From equation (\ref{theta}) it follows that, physically, $\theta$ depends
on the bulk sink strength density measured in units of the inverse time scale
for relaxation at the boundary.

\subsection{Sources and estimates of simulation inaccuracy}
\begin{figure}[t]
\begin{center}
   \epsfig{file=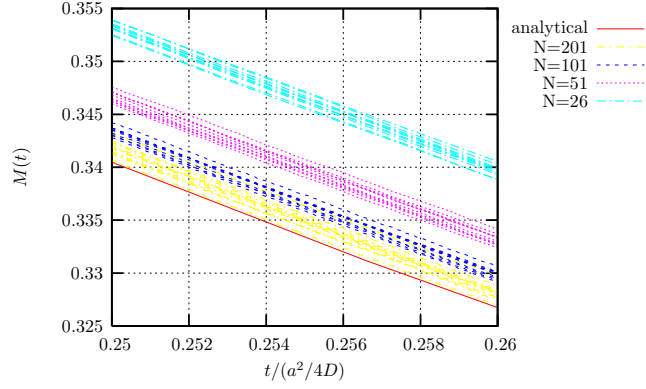,width=9cm}
\caption{$(N,\zeta)$-dependency for $\gamma= 1$, $10^6$ walkers
  (Mersenne Twister). Note that as $N$ increases the numerical
  solution moves towards the analytical solution}
\label{fig:nzeta}
\end{center}
\end{figure}
Figure \ref{fig:nzeta} shows the results of simulations of the average
magnetization $M$ for $\gamma=1$, in an interval of scaled time $t/\tau_D$.
The $N$ values 26, 51, 101 and 201 were used;
for each, 10 simulations with different random number generator seeds have
been plotted,
with $N_w=10^6$ random walkers released on the lattice in each simulation.
The lower most line shows the analytical solution in the same interval,
plotted as the sum of 50 modes.
The 'Mersenne Twister' generator\cite{mt} was used in this plot
as random number generator,
but also runs with single-precision versions of ran2 and
ran3\cite{nr} were made.
The simulations organize themselves in bands corresponding to the four $N$
values used, with $N$ increasing from above (the bands for $N=101$ and $N=201$
partially merging).

Various sources of error in the simulations can be discerned:
\begin{itemize}
\item Random:
\begin{itemize}
\item Small fluctuations in the curves increase with increasing $N$, for
      the same number of random walkers.
\item Generator seed / number of walkers: With each curve showing an average
      over $N_w=10^6$ walkers,
      for each $N$ value there is still a seed-dependent
      spread within a band of the order of $0.7\:\%$.
\item Truncation errors do not give important random effects in these
      simulations;
      curves for a single-precision ran3 (not shown) have about the same
      variance as those shown here.
\end{itemize}
\item Systematic:
\begin{itemize}
\item Finite $\epsilon$ effects: Since $\epsilon\propto1/(N-1)$, choosing $N$
      too small makes the order $\epsilon$ terms in equation (\ref{likn})
      increase in importance, thus violating equation (\ref{boundrw}) in addition to
      introducing inaccuracies in the representation of the derivatives in
      equation (\ref{diffrw}) and (\ref{boundrw}). The simulation results averaged over
      seed values show a deviation from the analytical solution as a
      monotonous function of $N$, of order $1\:\%$ for $N=101$ and rapidly
      increasing as $N$ decreases. (Technically, these $N$-dependent
      systematic deviations may be considered a 'scaling violation'.)
\item Truncation errors: For $N=26$ the results for ran3 (not shown) and for
      Mersenne Twister are very close, but for increasing $N$ the average difference
      between the bands becomes larger for ran3 than for Mersenne Twister, so that for
      large $N$ the bands actually make an 'undershoot' (not shown) of the
      analytical result. Such effects may arise from truncation when a random
      generator is used to make a choice with a given probability.
\end{itemize}
\end{itemize}
The level of accuracy may vary with $\gamma$ and with the range of scaled
time used. For the values treated above, we conclude that
$N\stackrel{>}{\sim}200$ and $N_w\stackrel{>}{\sim}10^6$ should be used to
obtain an accuracy in one given run of order $1\:\%$ or better.


\end{widetext}

\begin{thebibliography}{00}
\bibitem{song} Song Y.: Detection of the High Eigenmodes of Spin
  Diffusion in Porous Media. {\it Phys. Rev. Lett.} {\bf 85}, 3878 (2000)\\
  Song Y., Ryu S.,Sen P. N.: Determining multiple length scales in
  rocks. {\it Nature} {\bf 406}, 178 (2000)\\
  Song Y.: Pore sizes and pore connectivity in rocks using the effect of
internal field. {\it Mag. Reson. Img.} {\bf 19}, 417 (2001)\\
 Lisitza, M., Song Y.: Manipulation of the diffusion eigenmodes in
  porous media. {\it Phys. Rev. B} {\bf 65}, 172406 (2002)
\bibitem{freedman} Freedmann R., Heaton N., Flaum M., Hirasaki G.,
  Flaum C., H\"{u}rlimann M.: Wettability, Saturation, and Viscosity Using the
  Magnetic Resonance Fluid Characterizing Method and New
  Diffusion-Editing Pulse Sequences. {\it SPE} {\bf 77397},2002 SPE Annual
  Technical Conference and Ehibition, San Antonio, September 29-October 2.
\bibitem{heaton} Heaton N., Freedman R., Karmonik C., Taherian R., 
Walter K., Depavia L.: Application of a New-Generation NMR Wireline
  Logging Tool, {\it SPE} {\bf 77400}, 2002 SPE Annual Technical
  Conference and Exhibition, San Antonio, September 29 - October 2.
\bibitem{hurlimann} H\"{u}rlimann M., Venkataramanan L., Flaum C.,
 Speier P., Karmonik C., Freedman R., Heaton N.: Diffusion-Editing: New
 NMR Measurement of Saturation and Pore Geometry. The 2002 SPWLA
 Annual Logging Symposium, Houston, June 2-5.
\bibitem{hurlimann2} H\"{u}rliman M., Venkataramanan L.: Quantitative
  Measurement of Two-Dimensional Distribution Functions
of Diffusion and Relaxation in Grossly Inhomogeneous Fields.
{\it J. Mag. Reson.} {\bf 157}, 31-42, (2002)
\bibitem{brownstein1} Brownstein K., Tarr C.: Spin-Lattice Relaxation
  in a System Governed by Diffusion. {\it J. Mag. Reson} {\bf
  26}, 17-24 (1977)
\bibitem{brownstein2} Brownstein K., Tarr C.: Importance of Classical
  Diffusion in NMR Studies of Water in Biological Cells. {\it Phys.
  Rev.} {\bf A}(19) 2446-2453 (1979)
\bibitem{howard} J.J. Howard.: Quantitaive Estimates of Porous Media Wettability from
Proton NMR Measurments, {\it J. Mag. Reson. Img.} {\bf 16} 529-533 (1998)
\bibitem{zhang} Zhang Q., Huang C., Hirasaki G.: Interpretation of
Wettability in Sandstones With NMR Analysis, {\it Petrophysics}
{\bf 41}(3), 223-233 (2000)
\bibitem{radke} Radke C., Kovscek A., Wong H.: A Pore-Level Scenario for
  the Development of Mixed Wettability. {\it SPE} {\bf 24880},
  the 1992 Annual Technical Conference and Exhibition of the Society of
  Petroleum Engineers, Washington DC, October 4-7.
\bibitem{morrow1}Mason G., Morrow N.: Capillary behavior of a perfectly wetting
  liquid in irregular triangular tubes. {\it J. Coll. Int. Sci.} {\bf
  141}, 262-274 (1991)
\bibitem{morrow2} Morrow N., Mason G.: Recovery of Oil by Spontaneous
  Imbibition. {\it Curr Opin Coll Int Sci} {\bf 6}, 321 (2001)
\bibitem{blunt} Jackson M., Valvatne P., Blunt M.:
Prediction of wettability variation and its impact on waterflooding using pore- to
reservoir-scale simulation. {\it SPE} {\bf 77543}, 2002 SPE Annual Technical
  Conference and Exhibition, San Antonio, September 29 - October 2.\\ 
Piri M., Blunt M.: Pore-scale modeling of three-phase flow in
mixed-wet system. {\it SPE} {\bf 77726}, 2002 SPE Annual Technical
  Conference and Exhibition, San Antonio, September 29 - October 2.\\ 
Hui M., Blunt M.: Effects of Wettability on Three-Phase Flow in Porous
Media. {\it J. Phys. Chem.} {\bf B}(104), 3833-3845 (2000)
\bibitem{sorbie} van Dijke M., Sorbie K.: Three-phase capillary entry
  conditions in pores of noncircular cross-section. {\it J. Colloid
  Interface Sci} {\bf 260}, 385-397 (2003)
\bibitem{helland} Helland J., Skj\ae veland S. M.: Physically-Based
  Capillary Pressure  Correlation for Mixed Wet Reservoirs From
a Bundle of Tubes Model. {\it SPE} {\bf  89428}, 2004 SPE/DOE
  Fourteenth Symposium on Improved Oil Recovery, Tulsa, April 17-21.\\
Helland J., Skj\ae veland S. M.: Three-phase mixed-wet capillary pressure
  curves from a bundle-of-triangular-tubes model. The 8th
  International Symposium on Reservoir Wettability, Houston, May
  16-18, 2004.
\bibitem{al-mahrooqi} Al-Mahrooqi S., Grattoni C., Muggeridge A.,
  Zimmerman R., Jing X.: Pore-Scale Modelling of NMR Relaxation for the
  Characterization of Wettability. Paper presented at the 8th
  International Symposium on Reservoir Wettability, Houston, May 16-18,
  2004.
\bibitem{marinelli} Marinelli L., H\"{u}rlimann M. D. , Sen P. N.:
  Modal analysis of q-space relaxation correlation experiments.
  {\it J. Chem. Phys} {\bf 118}(19),8927-8940  (2003)
\bibitem{mccartin1} Mccartin B.: Eigenstructure of the Equilateral
  Triangle, Part II: The Neuman Problem  Triangle. {\it Math. Probl.
  Eng.} 
{\bf 8}, 517-539  (2002)
\bibitem{mccartin2} Mccartin B.: Eigenstructure of the Equilateral
  Triangle, Part I: The Dirichlet Problem, {\it SIAM Review} {\bf 45},  267-287
  (2003)
\bibitem{mccartin3} Mccartin B.: Eigenstructure of the Equilateral
  Triangle. Part III. The Robin Problem. {\it IJMMS} {\bf 16}, 807-825
  (2004)
%
\bibitem{rwk} Wilkinson D., Johnson D., Schwartz L.: Nuclear
  magnetic relaxation in porous media: The role of the mean lifetime
  $\tau(\rho,D)$. {\it Phys Rev B} {\bf 44},  4960-4973 (1991)\\ 
  Mitra P., Sen P., Schwartz L.: Short time behavior of
  the diffusion coefficient as a geometrical probe of porous media
  pore geometries. {\it Phys. Rev. B} {\bf 47}, 8565-8574 (1993) \\
  Mendelson K.: Continuum and random-walk models of magnetic
  relaxation in porous media, {\it Phys. Rev. B} {\bf 47}, 1081-1083 (1993)
  \\ 
   Mendelson K.: Percolation model of nuclear magnetic relaxation in porous media
, {\it Phys. Rev. B} {\bf 41},  562-567 (1990)\\ 
  Bergman D., Dunn K., Schwartz L., Mitra P.:
  Self-diffusion in a periodic porous medium: A comparison of
  different approaches, {\it Phys. Rev. E} {\bf 51}, 3393-3400 (1990) \\
  Toumelin E., Torres-Verdin C., Chen S.: Modeling of Multiple  
 Echo-Time NMR Measurements for Complex Pore
  Geometries and Multiphase Saturations.  {\it SPE} {\bf 85635},
  2002 SPE Annual Technical
  Conference and Exhibition, San Antonio, September 29 - October 2.
\bibitem{mt} Matsumoto M., Nishimura T.: Mersenne Twister: A 623-dimensionally
equidistributed uniform pseudorandom number generator, {\it ACM Trans.
Mod. and Comp. Sim.} {\bf 8}, 3-30 (1998) \\ 
Matsumoto M., Nishimura T.: Dynamic Creation of Pseudorandom
Number Generators, Monte Carlo and Quasi-Monte Carlo Methods 1998,
Springer, 2000, 56-69
\bibitem{nr} Press W. et al.: Numerical Recipes in C (Cambridge,
  1992).
\bibitem{vf1} Valfouskaya A., Adler P. M., Thovert J. -F.,
  Fleury M.: Nuclear-Magnetic-Resonance Diffusion Simulations in
  Porous Media {\it J. App. Phys.} {\bf 97} 083510 (2005) 
\bibitem{vf2} Valfouskaya A., Adler P. M.: Nuclear-Magnetic-Resonance Diffusion
  Simulations in Two Phases in Porous Media, {\it Phys. Rev. E} {\bf
  72} 056317 (2005) 
\bibitem{vf3} Valfouskaya A., Adler P. M., Thovert  J.-F., Fleury M.: 
Nuclear Magnetic Resonance Diffusion with Surface Relaxation
in Porous Media {\it J. Coll. and Int. Sci.} {\bf  295} 188 (2006)
\bibitem{zielenski} Zielinski L. J., Song Y., Ryu S., Sen P. N.: 
Characterization of coupled pore systems from the diffusion
eigenspectrum. {\it J. Chem. Phys.} {\bf 117}, 5361 (2002)
\bibitem{cpgm} Carr H., Purcell E.: Effects of diffusion on free precession in
NMR experiments, {\it Phys. Rev.} {\bf 94}, 630 (1954) \\ 
 Meiboom S., Gill D.: Compensation for pulse imperfections in
 Carr-Purcell NMR experiments. {\it Rev. Sci. Instrum.} {\bf 29}, 688
 (1958)
\bibitem{cal1} Callaghan P.: Principles of Nuclear Magnetic
  Resonance Microscopy.  Oxford Univ. Press, Oxford, 1991
\bibitem{price1} Price W.: Pulsed-field gradient nuclear magnetic
  resonance as a tool for studying translational diffusion: Part 1.
  Basic theory.{\it Concepts Magn. Reson.} {\bf 9}, 299-336 (1997)
\bibitem{price2} Price W.: Pulsed-field gradient nuclear magnetic
  resonance  as a tool for studying translational diffusion: Part II.
  Experimental aspects. {\it Concepts Magn. Reson.} {\bf 10}, 197-237 (1998)

\end{thebibliography}
\end{document}